\documentclass[a4paper]{article}

\usepackage[english]{babel}
\usepackage[utf8x]{inputenc}
\usepackage[T1]{fontenc}
\usepackage{booktabs}

\usepackage[a4paper,top=3cm,bottom=2cm,left=2.5cm,right=2.5cm,marginparwidth=1.75cm]{geometry}

\usepackage{amsmath}
\usepackage{graphicx}
\usepackage{tabularx}
\usepackage{algorithm}
\usepackage{algpseudocode}
\usepackage{multirow}
\usepackage{caption}
\usepackage{verbatim}
\usepackage{amsfonts}
\usepackage[colorinlistoftodos]{todonotes}
\usepackage[colorlinks=true, allcolors=blue]{hyperref}

\usepackage{orcidlink}

\definecolor{darkorchid}{rgb}{0.6, 0.2, 0.8}

\title{Trajectory-informed graph-based clustering for longitudinal cancer subtyping}
\author{Lara Cavinato$^{1}$\orcidlink{0000-0001-7007-0205}, Marco Rocchi$^{1}$, Luca Viganò$^{2,3}$\orcidlink{0000-0002-4108-4832}, Francesca Ieva$^{1,4}$\orcidlink{0000-0003-0165-1983}
\\
\small{$^1$MOX, Department of Mathematics, Politecnico di Milano, Milan, Italy} \\
\small{$^2$Hepatobiliary Unit, Department of Minimally Invasive General \& Oncologic Surgery,} \\ \small{Humanitas Gavazzeni University Hospital, Bergamo, Italy} \\
\small{$^3$Department of Biomedical Sciences, Humanitas University, Milan, Italy} \\
\small{$^4$Health Data Science Centre, Human Technopole, Milan, Italy}}
\date{}

\begin{document}
\maketitle

\begin{abstract}
Cancer subtyping plays a crucial role in informing prognosis and guiding personalized treatment strategies. However, conventional subtyping approaches often rely on static, biopsy-derived scores that hardly capture the biological heterogeneity and temporal evolution of the disease. In this study, we propose a novel trajectory-informed clustering method for cancer subtyping that integrates multi-modal clinical data and longitudinal patient trajectories. Our method constructs a patient similarity graph using time-varying imaging-derived features, clinical covariates, and transitions among key clinical states such as therapy, surveillance, relapse, and death. This graph structure enables the identification of patient subgroups that are not only phenotypically and genotypically distinct but also aligned with patterns of disease progression. We position our approach within the landscape of existing subtyping methods 
and highlight its advantages in terms of temporal modeling and graph-based interpretability. Through simulation studies and application to a real world dataset of liver metastases, we demonstrate the ability of our framework to uncover clinically relevant subtypes with distinct prognostic trajectories. Our results underscore the potential of trajectory-informed clustering to enhance personalized oncology by bridging cross-sectional biomarkers with dynamic disease evolution. An R implementation can be found at this \href{http://www.github.com/MarcoRocchi/MS3C}{link}.
\end{abstract}

\section{Introduction}

Cancer remains one of the leading causes of mortality worldwide, accounting for nearly 10 million deaths in 2020 alone, according to the World Health Organization \cite{cancer_who}. Despite decades of research and significant advancements in diagnostics and therapeutics, cancer continues to present difficult challenges to clinicians and researchers. These challenges stem not only from the aggressive nature of many malignancies but more fundamentally from the intrinsic complexity and heterogeneity of the disease across individuals, subtypes, and stages \cite{dagogo2018tumour}.

The term cancer encompasses a wide family of diseases characterized by uncontrolled cellular proliferation, but the clinical and molecular presentation of each case can vary dramatically. This heterogeneity manifests across several dimensions: from genetic mutations and epigenetic alterations at the molecular level, to tumour microenvironment dynamics, patient immune response, and resistance to therapy. Moreover, tumours can evolve under selective pressures, such as treatment, leading to further divergence in disease progression and clinical outcome. These layers of complexity make cancer a moving target, which in turn complicates the development of universally effective therapies and stratification systems \cite{simpson2025challenges}.

Despite the biological diversity of cancer, clinical treatment guidelines are often built upon a limited number of criteria. Standardized approaches typically rely on tumour histology, anatomical site, and coarse staging information derived from diagnostic biopsies. These indicators, while clinically valuable, fail to fully represent the variability of individual cases \cite{hodel2018impact}. Personalized treatment decisions based on molecular and phenotypic characteristics, both at the microscale (e.g., genomics, transcriptomics, proteomics) and macroscale (e.g., imaging phenotypes, patient-reported symptoms), remain largely underutilized in clinical protocols. As a result, the current paradigm risks under-treating aggressive forms or over-treating mild cases, hampering the efficacy and safety of interventions \cite{yang2022personalized}.

Cancer subtyping, the classification of tumours into biologically and clinically relevant and prognostic subgroups, is a growing field that aims to address these limitations \cite{duan2021evaluation}. Traditional subtyping systems often rely on static, population-level biopsy scores that are limited in scope and fail to capture the dynamic nature of disease progression. These systems also neglect longitudinal aspects of cancer evolution, such as response to therapy, relapse, and survival outcomes. In contrast, modern subtyping approaches aim to integrate a broad range of data sources, including time-resolved imaging, clinical covariates, and high-dimensional omics data, to construct refined patient stratifications that are both more representative and potentially actionable \cite{zhao2019molecular, wang2023precision}.
To this end, several recent methodologies have emerged for cancer subtyping based on multi-source data and longitudinal observations. Profile regression, for instance, is a Bayesian clustering framework that simultaneously models covariates and outcomes using a Dirichlet process mixture model \cite{liverani2021clustering}. It provides flexible clustering while accounting for uncertainty, but it can be computationally intensive and less effective in high-dimensional settings. Nearest shrunken centroid methods, as used in gene expression analysis, classify samples by shrinking class centroids towards the overall centroid to improve discrimination \cite{bair2004semi}. While simple and interpretable, these methods are typically limited to cross-sectional data and may not capture temporal dynamics. Variational deep survival clustering combines deep learning representations with survival analysis to learn patient subgroups with distinct prognostic patterns \cite{manduchideep}. Although powerful, this approach requires careful tuning and is often a black-box. Graph-based survival clustering methods, in contrast, model patient similarities as a network, using graph structures to incorporate both feature similarity and survival behavior \cite{liu2020supervised}. This class of methods excels at capturing complex dependencies and relational structures among patients, although estimating the underlying graph from noisy, high-dimensional data remains challenging.

In this work, we propose a novel graph-based clustering method specifically designed for cancer subtyping that integrates both temporal and cross-sectional patient information. Our approach constructs a patient similarity graph informed by time-varying imaging-derived features, clinical covariates, and the sequence of transitions among critical clinical states such as therapy, surveillance, relapse, and death. These transitions are modeled as trajectories that capture the evolution of the disease, enabling us to group patients not only by static characteristics but also by their disease course. By encoding temporal information directly into the graph structure, our method facilitates the identification of clinically relevant subtypes that reflect both phenotypic and genotypic diversity, while also being sensitive to disease dynamics.

The main contribution of this paper is the development of this trajectory-informed, graph-based clustering framework for robust and interpretable cancer subtyping. In contrast to existing approaches, our model leverages longitudinal clinical trajectories to inform the graph structure, thereby enabling subtype definitions that are sensitive to both patient-specific evolution and multimodal data integration. The long term goal is to derive imaging, possibly time-varying, biomarkers that correlate with tumour evolution to be used to guide clinical decision making.

The remainder of this paper is structured as follows. In Section \ref{sec:model-formulation}, we present the formal model formulation and describe how the patient similarity graph is constructed from multimodal and temporal data. Section \ref{sec:model-optimization} details the optimization procedure for clustering and inference. Section \ref{sec:simulation-study} evaluates the performance of our method through extensive simulation studies designed to test sensitivity to noise, missing data, and temporal resolution. Section \ref{sec:application} demonstrates the utility of our framework in a real-world case study on liver metastases, highlighting the clinical and biological relevance of the derived subtypes. Finally, Section \ref{sec:conclusions} concludes with a discussion of future directions and potential clinical applications.

\section{Model formulation}
\label{sec:model-formulation}

In this section, we present a framework to model patient similarity and subtyping by jointly leveraging static covariates and time-to-event information from multiple transitions, modeled through a Multi-State Model. The approach builds a similarity graph among patients, where the similarity score is not precomputed but learned from data, guided by survival outcomes. This results in a data-driven affinity matrix that better captures disease trajectories and patient phenotypes.

\subsection{Preliminaries on Multi-State Models}

Multi-State Models (MSM) are an extension of survival analysis used to describe processes in which individuals move through a finite number of states over time \cite{putter2007tutorial}. These models are particularly well-suited for applications in clinical research, where patients may experience multiple events or transitions, such as disease progression, remission, or death. The key idea is to model the stochastic process \( \{X(t), t \geq 0\} \), where \( X(t) \) denotes the state occupied by an individual at time \( t \).

\begin{figure}
    \centering
    \includegraphics[width=0.9\linewidth]{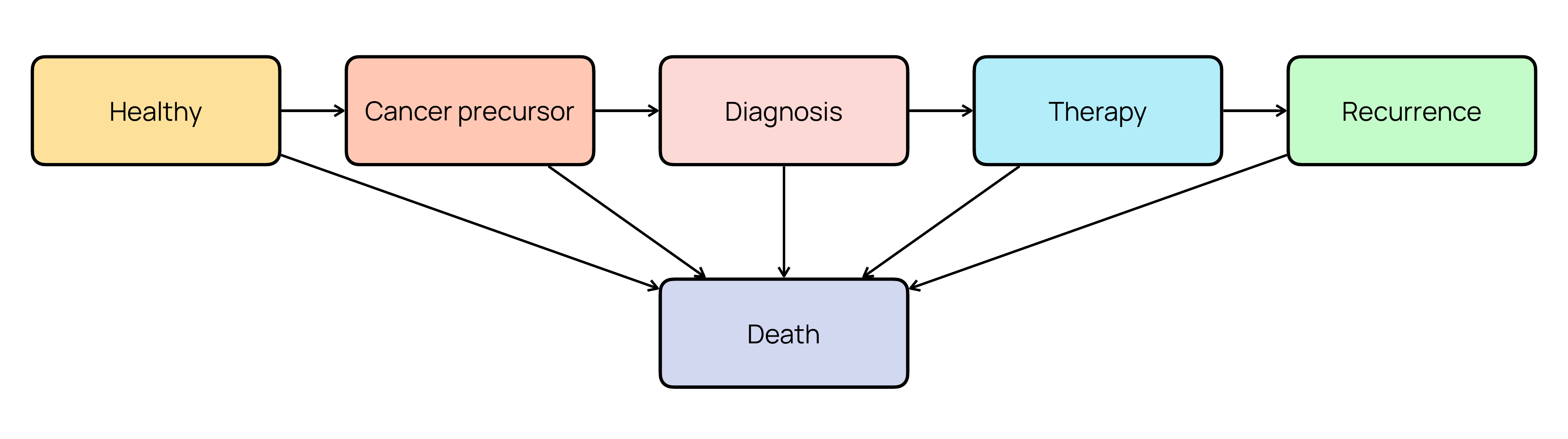}
    \caption{Example of a clock-reset multi-state model (MSM) representing the natural history and clinical management of cancer. The states reflect key phases of disease progression: \textit{Healthy} represents the absence of malignancy; \textit{Cancer precursor} indicates a subclinical or pre-malignant condition; \textit{Diagnosis} marks the point of clinical detection; \textit{Therapy} includes the treatment phase such as surgery, chemotherapy, or radiotherapy; \textit{Recurrence} denotes relapse after treatment; and \textit{Death} is the absorbing terminal state. Arrows represent possible transitions between states, and time is reset at each state transition, consistent with a clock-reset MSM framework. This structure allows for modelling heterogeneous disease trajectories and estimating transition hazards between clinically relevant states.}
    \label{fig:msm-example}
\end{figure}

Let \( \mathcal{S} = \{1, 2, \ldots, K\} \) denote the finite set of possible states. The process starts in an initial state (e.g., \( X(0) = 1 \)) and evolves over time as transitions occur between states. Transitions can be either \textit{reversible}, where individuals can return to a previously visited state, or \textit{irreversible}, where once a transition occurs, the individual cannot return (e.g., progression to death).
The structure of allowed transitions is often visualized as a directed graph, where nodes represent states and edges represent possible transitions. An example of MSM in the context of cancer clinical pathway is given in Figure \ref{fig:msm-example}.

The fundamental quantities of interest in a multi-state model are the transition intensities or \textit{hazards}:

\[
\lambda_{rs}(t) = \lim_{\Delta t \to 0} \frac{P(X(t + \Delta t) = s \mid X(t) = r)}{\Delta t}, \quad r \neq s
\]

which define the instantaneous risk of moving from state \( r \) to state \( s \) at time \( t \). A common assumption is that the process follows the \textit{Markov property}, meaning that future transitions depend only on the current state and not on the past history, i.e.:

\[
P(X(t + \Delta t) = s \mid \mathcal{F}_t) = P(X(t + \Delta t) = s \mid X(t) = r)
\]

where \( \mathcal{F}_t \) denotes the filtration (history) up to time \( t \).

In a semi-parametric setting, the Cox proportional hazards model can be used to model the transition intensities. For a transition from state \( r \) to state \( s \), the hazard function for individual \( i \) at time \( t \) is given by:

\[
\lambda_{rs}^{(i)}(t) = \lambda_{0,rs}(t) \exp\left( \beta_{rs}^T \mathbf{X}_i^{rs}(t) \right)
\]

Here, \( \lambda_{0,rs}(t) \) is the unspecified baseline hazard for the \( r \rightarrow s \) transition, \( \beta_{rs} \) is a vector of regression coefficients, and \( \mathbf{X}_i^{rs}(t) \) is the vector of covariates (possibly time-dependent) relevant to the transition.

Inference in multi-state models is typically based on partial likelihoods, constructed from observed transition times and states. Let \( T_i \) denote the observed transition time and \( \delta_i \) the indicator of the observed transition (from state \( r \) to \( s \)). The partial likelihood for the transition \( r \rightarrow s \) across individuals is given by:

\[
L_{rs}(\beta_{rs}) = \prod_{i \in \mathcal{D}_{rs}} \frac{\exp\left( \beta_{rs}^T \mathbf{X}_i^{rs}(T_i) \right)}{\sum_{j \in \mathcal{R}_{rs}(T_i)} \exp\left( \beta_{rs}^T \mathbf{X}_j^{rs}(T_i) \right)}
\]

where \( \mathcal{D}_{rs} \) is the set of individuals who experience the transition \( r \rightarrow s \), and \( \mathcal{R}_{rs}(T_i) \) is the risk set at time \( T_i \).

Multi-state models are particularly useful in settings where patients may experience intermediate events before a terminal event, such as disease progression through well-defined clinical stages, relapse and recovery in chronic diseases, time to treatment and adverse events in pharmacological studies.
They provide a more nuanced understanding of the disease process than traditional survival models, enabling researchers to assess transition-specific risks and the effect of covariates on different parts of the clinical pathway.

\subsection{Population Graph Estimation}

As stated in the introduction, we aim to learn a graph structure describing the population under study in a way that allows us to directly segment such graph is groups of patients displaying both similar clinical representations and similar histories. To do this, we exploit graph and MSM literature.

We define a similarity graph \( G = (V, E) \), where each node \( v_i \in V \) corresponds to a patient, and each edge \( (v_i, v_j) \in E \) encodes the similarity between patients \( i \) and \( j \). This graph is represented by an affinity matrix \( S \in \mathbb{R}^{n \times n} \), where each entry \( S_{ij} \) quantifies how similar patient \( i \) is to patient \( j \). A higher value of \( S_{ij} \) indicates greater similarity between the two patients.

Rather than computing this matrix using unsupervised or heuristic rules, we propose to learn \( S \) in a supervised fashion, using both the covariate profiles and disease progression trajectories of each patient, as informed by their observed survival transitions.

\subsubsection{Learning the Graph Similarity Matrix via Distance Minimization}

Let \( d(i, j) \) be a distance function that captures the dissimilarity between patient \( i \) and patient \( j \). Our objective is to assign higher similarity scores to pairs of patients that are more similar according to this distance. To formalize this intuition, we aim to minimize the total distance between similar pairs in the graph by solving the following optimization problem \cite{nie2014clustering}:

\begin{equation}
\label{eq:similarity_unreg}
\min_{S} \sum_{i=1}^n \sum_{j=1}^n d(i, j) S_{ij}
\quad \text{subject to} \quad
\sum_{j=1}^n S_{ij} = 1,\quad S_{ij} \ge 0\ \forall i,j.
\end{equation}

This formulation enforces that the similarity scores associated with each patient sum to one, and that all similarity values are non-negative. As a result, each row of the matrix \( S \) can be interpreted as a distribution over neighbors, assigning probabilistic similarity weights to other patients.

However, this formulation can lead to degenerate solutions, such as the case in which each patient assigns all its similarity mass to a single nearest neighbor. To prevent such collapse, we introduce a regularization term that penalizes highly concentrated similarity vectors \cite{nie2014clustering}:

\begin{equation}
\label{eq:similarity_reg}
\min_{S} \sum_{i=1}^n \sum_{j=1}^n \left( d(i, j) S_{ij} + \lambda S_{ij}^2 \right)
\quad \text{subject to} \quad
\sum_{j=1}^n S_{ij} = 1,\quad S_{ij} \ge 0.
\end{equation}

This regularized formulation encourages more balanced similarity distributions, with the parameter \( \lambda \geq 0 \) controlling the degree of smoothness.

The definition of the distance function \( d(i, j) \) plays a central role in learning meaningful similarities. We propose a function that combines two complementary components: one based on baseline covariates, and another based on patient-specific risk scores derived from a multi-state survival model. The combined distance is written as:

\begin{equation}
\label{eq:distance_decomposition}
d(i,j) = \mu\, d_{\text{cov}}(i,j) + d_{\text{msm}}(i,j),
\end{equation}

where \( \mu \geq 0 \) is a scalar hyperparameter that balances the contribution of the two components.

\paragraph{Covariate-based Distance}

The first term \( d_{\text{cov}}(i,j) \) captures differences in baseline patient characteristics. Each patient \( i \) is associated with a vector \( \mathbf{X}_i \in \mathbb{R}^p \) with $p$ number of unique covariates across all transitions (i.e., covariates such as sex or age, which may appear in multiple transitions, are included only once). The covariate-based distance is then computed as the squared Euclidean distance between these vectors:

\begin{equation}
\label{eq:covariate_distance}
d_{\text{cov}}(i,j) = \|\mathbf{X}_i - \mathbf{X}_j\|_2^2.
\end{equation}

\paragraph{MSM-based Distance}

The second term, \( d_{\text{msm}}(i,j) \), measures the difference in the predicted risk trajectories of the two patients, using a Multi-State Model. 

We assume that patients follow specific clinical pathways defined by \( K \) state-to-state transitions. For each transition \( k \), patients are described by a set of covariates, denoted as \( \mathbf{X}^{k} \), which may differ from the covariates describing other transitions. 
These covariates can represent various clinical and demographic factors, such as age, gender, clinical markers, as well as other relevant biomarkers derived from both invasive and non-invasive disease assessments. For instance, in our case study, we exploit time-resolved imaging data to extract useful patterns in the dynamics of imaging biomarkers, from a tumour texture point of view. In fact, imaging has been proven to entail useful clinical information while being non-invasively acquired during clinical routine.
The transitions between these states can be modeled using stratified Cox models, which describe the hazard of transitioning from one state to another as a function of the covariates.

To define a distance between two patients, say \( i \) and \( j \), who belong to the set of patients following this clinical pathway, we aim to calculate a weighted distance that reflects the differences in the Cox-related risks for transitioning between states. The approach involves computing the difference in transition hazards, where the risks of transitioning depend on the covariates of the patients, and then aggregating these differences in a weighted way across all possible transitions to quantify the similarity between trajectories.

The stratified Cox model provides a framework to model the hazard for transitioning between states. The hazard function \( \lambda_k(t | \mathbf{X}^k) \) for a patient in the transition \( k \) at time \( t \) is given by the following expression:

\[
\lambda_k(t | \mathbf{X}^k) = \lambda_0(t) \exp(\beta_k^T \mathbf{X}^k)
\]

Here, \( \lambda_0(t) \) represents the baseline hazard function, which is common across all patients, while \( \beta_k^T \mathbf{X}^k \) is the linear predictor for the transition \( k \) hazard from state. The vector \( \beta_k \) contains the regression coefficients associated with the covariates \( \mathbf{X}^k \) of the patient, and the exponential term reflects how these covariates influence the risk of transitioning between states.

The difference in the risk between two patients \( i \) and \( j \) for a given transition \( k \) can be measured using the \textit{hazard ratio} (HR), which is defined as the ratio of their hazards for that transition. The hazard ratio for the transition \( k \) between patients \( i \) and \( j \) is given by:

\[
\text{HR}_{i,j}^k = \frac{\lambda_k(t | \mathbf{X}_i^k)}{\lambda_k(t | \mathbf{X}_j^k)} = \frac{\lambda_0(t)\exp(\beta_k^T \mathbf{X}_i^k)}{\lambda_0(t)\exp(\beta_k^T \mathbf{X}_j^k)} = \frac{\exp(\beta_k^T \mathbf{X}_i^k)}{\exp(\beta_k^T \mathbf{X}_j^k)} = \exp\left(\beta_k^T (\mathbf{X}_i^k - \mathbf{X}_j^k)\right)
\]

This hazard ratio reflects how much the risk of transitioning differs between patients \( i \) and \( j \). A hazard ratio greater than 1 indicates that patient \( i \) has a higher risk of transitioning compared to patient \( j \), while a hazard ratio less than 1 suggests the opposite.

To define a distance between patients \( i \) and \( j \) based on their clinical pathways, we need to aggregate the differences in their risks for transitioning between all \( K \) state-to-state transitions. The key idea is to compute the \textit{logarithmic difference} in hazard ratios for each transition and then weight these differences according to the precision with which the Cox model estimates the corresponding hazard \cite{scheike2007direct}. This precision can be captured by the inverse of the variance of the regression coefficients \( \beta_k \).

Let the weighted distance \( d_{msm}(i,j) \) between patients \( i \) and \( j \) be defined as the sum of the weighted differences across all states. The difference for each transition is proportional to the absolute difference in the linear predictor \( \beta_k^T (\mathbf{X}_i^k - \mathbf{X}_j^k) \), with the weights reflecting the uncertainty in the estimates of the transition hazards. A natural form for this squared weighted distance can be expressed as follows. We take the squared form for compliance with \( d_{cov}(i,j) \) distance:

\[
d_{msm}(i,j) = \sum_{k=1}^K w_k \cdot \left\| \beta_k^T (\mathbf{X}_i^k - \mathbf{X}_j^k) \right\|_2^2
\]

Here, \( K \) denotes the total number of transitions, and \( w_k \) represents the weight assigned to each of them. These weights can be designed in various ways, but a common approach is to set \( w_k \) to be inversely proportional to the estimated variance of the regression coefficients, \( \hat{\text{Var}}(\beta_k) \), reflecting the precision of the transition-specific Cox model estimates \cite{greenland1987quantitative, hastie2009elements, carlin2008bayesian}, as more uncertain estimates (larger variances) should contribute less to the overall distance. Thus, the weight for each transition would be:

\[
w_k = \frac{1}{\hat{\text{Var}}(\beta_k)}
\]

In this case, the distance \( d_{msm}(i,j) \) between two patients could be written as:

\[
d_{msm}(i,j) = \sum_{k=1}^K \frac{\left\| \beta_k^T (\mathbf{X}_i^k - \mathbf{X}_j^k) \right\|_2^2}{\hat{\text{Var}}(\beta_k)}
\]

In this formulation, the inverse variance \( \frac{1}{\hat{\text{Var}}(\beta_k)} \) serves as a weight that reflects the precision of the Cox model's estimate for each transition. Larger variances in \( \beta_k \) (indicating less precision) reduce the influence of the corresponding transition on the overall distance between the two patients \cite{marin2010weighting, fisher2024inverse}.
Alternatively, one could use other forms of weighting based on clinical relevance or empirical evidence regarding the importance of different transitions \cite{lin2021combining, scheike2007direct}.

The resulting distance measure provides a way to quantify how similar or different two patients are in terms of their clinical pathway. A smaller distance suggests that the two patients have similar risk profiles for transitioning between states, while a larger distance indicates that their risk profiles are more distinct. This weighted distance captures both the difference in covariates across states and the uncertainty associated with the estimation of transition hazards, making it a useful tool for comparing patients in a their clinical evolution, to be considered together with the mere distance between patients' covariates.

\subsubsection{Learning Transition Parameters via Stratified Cox Loss}

The MSM-based distance \( d_{msm}(i,j) \) introduced previously depends on the transition-specific regression coefficients \( \beta_k \), which must be estimated from data. To obtain these estimates, we employ a stratified Cox proportional hazards model, where each transition \( k \in \{1, \ldots, K\} \) is treated as a separate stratum with its own covariates and parameter vector \( \beta_k \).

Let \( \delta_i^{k} \) denote the event indicator for patient \( i \) associated with transition \( k \), such that \( \delta_i^{k} = 1 \) if patient \( i \) undergoes the transition \( k \), and \( \delta_i^{k} = 0 \) otherwise. Let \( \mathcal{R}_i^{k} \) be the risk set for patient \( i \) at the time of their observed event or censoring in stratum \( k \), and let \( \mathbf{X}_i^{k} \) be the covariate vector of patient \( i \) relevant to transition \( k \).

Being the partial likelihood for the transition \( k \) across individuals previously defined 

\[
L_{k}(\beta_{k}) = \prod_{i \in \mathcal{D}_{k}} \frac{\exp\left( \beta_{k}^T \mathbf{X}_i^{k}(T_i) \right)}{\sum_{j \in \mathcal{R}_{k}(T_i)} \exp\left( \beta_{k}^T \mathbf{X}_j^{k}(T_i) \right)},
\]

we derive the stratified partial likelihood across all transitions.

Under standard assumptions such as independence of transitions (given covariates), non-informative censoring, markovianity and separate risk sets and time scales for each transition, the overall likelihood of the multi-state process can be factorized and defined as the summation of log-partial-likelihoods across transitions as theoretically justified under the stratified Cox framework \cite{andersen1993statistical}:


\[
\mathcal{L}(\beta) = - \sum_{k=1}^{K} \sum_{i=1}^{n} \delta_i^{k} \left( \beta_k^T \mathbf{X}_i^{k} - \log \sum_{j \in \mathcal{R}_i^{k}} \exp\left( \beta_k^T \mathbf{X}_j^{k} \right) \right)
\]

This loss function estimates the transition-specific coefficients \( \beta_k \) by maximizing the stratified Cox partial likelihood, accounting for the heterogeneity across transitions. To promote sparsity in the learned coefficients and facilitate model interpretability and variable selection, we augment the loss with an \( \ell_1 \)-regularization term. The total regularized loss function is given by:

\[
\mathcal{L}_{\text{reg}}(\beta) = \mathcal{L}(\beta) + \eta \sum_{k=1}^{K} \|\beta_k\|_1
\]

where \( \eta \geq 0 \) is a hyperparameter that controls the degree of regularization. The optimization of \( \mathcal{L}_{\text{reg}}(\beta) \) yields the set of transition-specific parameter vectors \( \beta_k \) that are used in defining the patient-specific transition risks and, consequently, the distance \( d_{msm}(i,j) \) between individuals.

\subsubsection{Joint Objective Function}

Bringing together all the defined components into a joint objective function that simultaneously learns the parameters \( \{\beta_{k}\} \) and the similarity matrix \( S \). The final optimization problem is defined as:

\begin{equation}
\label{eq:final_loss}
\begin{aligned}
\min_{\beta, S} \quad & 
\underbrace{ - \sum_{k=1}^{K} \sum_{i=1}^{n} \delta_i^{k} \left( \beta_k^T \mathbf{X}_i^{k} - \log \sum_{j \in \mathcal{R}_i^{k}} \exp\left( \beta_k^T \mathbf{X}_j^{k} \right) \right) }_{\text{ Stratified Cox partial log-likelihood}} \\
& + \underbrace{ \eta \sum_{k=1}^{K} \|\beta_k\|_1 }_{\text{Sparsity via } \ell_1\text{-regularization}} \\
& + \underbrace{ \gamma \sum_{i=1}^n \sum_{j=1}^n \left( \mu \left\|\mathbf{X}_i - \mathbf{X}_j\right\|_2^2 + \sum_{k=1}^K w_k \cdot \left\| \beta_k^T (\mathbf{X}_i^k - \mathbf{X}_j^k) \right\|_2^2 \right) S_{ij} + \lambda S_{ij}^2 }_{\text{Similarity-regularized patient distance learning}} \\
\text{subject to} \quad & \sum_{j=1}^n S_{ij} = 1, \quad S_{ij} \ge 0 \quad \forall i, j.
\end{aligned}
\end{equation}

The first term in this objective corresponds to the stratified Cox partial log-likelihood, used to estimate the survival risk for each transition. The second term imposes sparsity through an \( \ell_1 \)-norm on the transition weights. The third term integrates the learned similarity scores, by minimizing a regularized and weighted distance between patients in both covariate and risk space. The hyperparameters \( \gamma \) and \( \mu \) control the influence of the similarity structure in the overall optimization.

This joint learning framework integrates survival modelling with similarity learning in a principled and interpretable manner. By combining a stratified Cox formulation with a graph-based regularized objective, the method captures both baseline differences and disease trajectory similarities, enabling nuanced patient stratification that is informed by both static and temporal data.

\subsection{Convergence analysis} To establish the theoretical soundness of the proposed joint optimization framework, we now examine its convergence properties. In particular, we aim to demonstrate that, under standard regularity conditions, the objective function defined in \eqref{eq:final_loss} admits a well-posed solution. The following proof shows that the optimization problem is convex in each block of variables when the others are fixed, that the subproblem in the similarity matrix $S$ is strictly convex with a unique minimizer, and that substituting this solution yields a reduced problem in $w$ that remains convex. These properties together ensure the existence and uniqueness of a global minimum for the joint problem, thus guaranteeing stability and convergence of the proposed learning algorithm.

\vspace{0.5em}
\textbf{Step 1: Reformulating the objective function. \\}
The optimization problem \eqref{eq:final_loss} can be rewritten as:
\begin{equation}
\min_{\beta,S} \; f(\beta) + \lambda \|S\|_2^2 + \gamma \langle g(\beta), S \rangle,
\end{equation}
Rewriting the quadratic term explicitly:
\begin{equation}
\min_{\beta,S} \; f(\beta) + \frac{\lambda}{2} S^\top I S + \langle \gamma g(\beta), S \rangle,
\end{equation}
where $f(\beta)$ is a function of $\beta$ (representing the original objective terms depending solely on $\beta$),  $\|S\|_2^2$ denotes the squared Frobenius norm of $S$, ensuring convexity, $\langle g(\beta), S \rangle$ is the inner product between $S$ and a function $g(\beta)$ and $\lambda > 0$ and $\gamma$ are scalar parameters.\\

\vspace{0.5em}
\textbf{Step 2: Convexity of the problem. \\}
The term $\frac{\lambda}{2} S^\top I S = \frac{\lambda}{2} \|S\|_2^2$ is strictly convex in $S$, while the linear term $\langle \gamma g(\beta), S \rangle$ is affine in $S$. Assuming $f(\beta)$ is convex in $\beta$, the constraints on $S$
\[
\sum_j S_{ij} = 1, \quad S_{ij} \ge 0, \quad \forall i,j,
\]
define a convex polytope and the optimization problem in $S$ for a given $\beta$ is a strictly convex quadratic program. Therefore, for any fixed $\beta$, the problem
\[
S^*(\beta) = \arg\min_S \frac{\lambda}{2} \|S\|_2^2 + \gamma \langle g(\beta), S \rangle
\]
admits a unique solution $S^*(\beta)$, since quadratic objectives with strictly positive-definite terms ($\lambda > 0$) have unique minimizers over convex sets.\\

\vspace{0.5em}
\textbf{Step 3: Reduced problem and existence of solution. \\}
Substituting $S^*(\beta)$ into the original problem yields the reduced problem:
\[
\min_\beta f^*(\beta) = f(\beta) + \lambda \|S^*(\beta)\|_2^2 + \gamma \langle g(\beta), S^*(\beta) \rangle
\]

Since $S^*(\beta)$ is uniquely defined for each $\beta$, $f^*(\beta)$ remains convex whenever
$f(\beta)$ is convex. Moreover, if $f(\beta)$ is strongly convex, then $f^*(\beta)$ is also
strongly convex, ensuring a unique global minimum with respect to $\beta$.\\

\vspace{0.5em}
\noindent
\textbf{Step 4: Compactness and global minimum. \\}
If $\beta$ is restricted to a compact domain (which is often the case in regularized
learning problems) then $f^*(\beta)$ attains its minimum within this domain. Because
$f^*(\beta)$ is convex (or strongly convex under additional conditions), the minimizer is unique.\\

\vspace{0.5em}
\noindent
The joint optimization problem therefore admits a unique pair $(\beta^*, S^*(\beta^*))$ corresponding to a global minimum.

\section{Model optimization}
\label{sec:model-optimization}

The affinity matrix estimated in the previous section needs to be estimated and exploited to adaptively perform clustering of nodes. The resulting segmentation of the graph resolves by construction into the subtyping of patients based on similarity in both their baseline and disease evolution characteristics. In this section, we explain how to solve the optimization problem.

\subsection{Population Graph Segmentation}

Starting from the optimization problem defined in Equation \ref{eq:final_loss}, we observe that the variable \( S \in \mathbb{R}^{n \times n} \) represents a similarity or assignment matrix, which defines the weighted edges of a graph whose vertices correspond to the data points \( \{\mathbf{X}_i\}_{i=1}^n \). In this graph, an edge between node $i$ and node $j$ has weight $S_{ij}$.
The goal is to impose an additional constraint or regularization on 
$S$ such that the resulting graph has exactly $c$ connected components, where $c$ is the desired number of clusters into which we wish to partition the data.
To achieve this, we exploit a fundamental property from spectral graph theory. Specifically, given the (unnormalized) graph Laplacian matrix $L$ defined as:

\begin{equation}
    L = D - \frac{S+S^T}{2}
\end{equation}

where $D$ is the degree matrix with diagonal entries

\begin{equation}
    D_{ii} = \sum_{j=1}^n \frac{S_{ij}+S_{ji}}{2}
\end{equation}

it is known that the multiplicity of the eigenvalue $0$ of $L$ is exactly equal to the number of connected components of the graph \cite{chung1997spectral}. In other words, if $L$ has $c$ eigenvalues equal to zero, then the graph has precisely $c$ disjoint connected components. Thus, in order to enforce that the graph constructed from $S$ has exactly $c$ connected components, we must ensure that $L$ has exactly $c$ eigenvalues equal to zero.

Directly enforcing an exact constraint on the spectrum of $L$ would be computationally prohibitive and non-convex. Instead, we can introduce an additional optimization objective that encourages $L$ to have $c$ eigenvalues close to zero. A classical way to approximate this idea is to introduce a matrix \( U \in \mathbb{R}^{n \times c} \) whose $i$-th row is the vector \(u_i \in \mathbb{R}^{1 \times c}\) assigned to each individual \(i\) defining their association strength to each cluster \(k\) and whose columns are constrained to be orthonormal, i.e.,

\begin{equation}
    U^{T}U = I_c
\end{equation}
 
and penalize the trace of the quadratic form $U^{T}LU$, leading to an additional regularization term of the form:

\begin{equation}
    \alpha \cdot \mathrm{Tr}(U^{T}LU)
\end{equation}

where $\alpha > 0$ is a hyperparameter controlling the strength of the spectral constraint.

The intuition behind this penalty is that minimizing $\mathrm{Tr}(U^{T}LU)$ over all matrices $U$ with orthonormal columns forces the columns of $U$ to align with the eigenvectors associated with the smallest eigenvalues of $L$. If the trace is minimized sufficiently, the corresponding eigenvalues are driven towards zero, thereby encouraging the emergence of $c$ connected components.
Consequently, the full optimization problem becomes:

\begin{equation}
\label{eq:trace_loss}
\begin{aligned}
\min_{\beta, S, U} \quad & 
- \sum_{k=1}^{K} \sum_{i=1}^{n} \delta_i^{k} \left( \beta_k^T \mathbf{X}_i^{k} - \log \sum_{j \in \mathcal{R}_i^{k}} \exp\left( \beta_k^T \mathbf{X}_j^{k} \right) \right) +\eta \sum_{k=1}^{K} \|\beta_k\|_1\\
& + \gamma \sum_{i=1}^n \sum_{j=1}^n \left( \mu \left\|\mathbf{X}_i - \mathbf{X}_j\right\|_2^2 + \sum_{k=1}^K w_k \cdot \left\| \beta_k^T (\mathbf{X}_i^k - \mathbf{X}_j^k) \right\|_2^2 \right) S_{ij} + \lambda S_{ij}^2\\
& + \alpha \cdot \mathrm{Tr}(U^{T}LU)\\
\text{subject to} \quad & \sum_{j=1}^n S_{ij} = 1, \quad S_{ij} \ge 0 \quad \forall i, j, \quad U^{T}U = I_c.
\end{aligned}
\end{equation}

Thus by augmenting the original optimization problem with the spectral regularization term $\alpha \cdot \mathrm{Tr}(U^{T}LU)$, and ensuring that $U$ has orthonormal columns, we effectively guide the learning process towards constructing a graph with exactly $c$ connected components. This process makes the neighborhood assignment $S$ an adaptive mechanism for clustering, fully integrated into the learning of the model parameters $\beta$.

\subsubsection{Intuitive Explanation of Minimizing $\mathbf{\mathrm{Tr}(U^{T}LU)}$}

Let us consider the Laplacian matrix \(L \in \mathbb{R}^{n \times n}\) associated with a graph. Since \(L\) is symmetric and positive semi-definite, it has a complete set of orthonormal eigenvectors \(\{v_1, v_2, \ldots, v_n\}\) and real non-negative eigenvalues \(0 \leq \lambda_1 \leq \lambda_2 \leq \cdots \leq \lambda_n\).

We are interested in solving the following minimization:
\[
\min_{U \in \mathbb{R}^{n \times c}} \quad \text{Tr}(U^T L U) \quad \text{subject to} \quad U^T U = I_c,
\]
where \(U\) is a matrix with \(c\) orthonormal columns. Express \(U\) in terms of the eigenvectors of \(L\):
\[
U = V Y,
\]
where \(V = [v_1, \ldots, v_n]\) is the orthogonal matrix of eigenvectors of \(L\), and \(Y \in \mathbb{R}^{n \times c}\) is a coefficient matrix. Since \(L = V \Lambda V^T\) with \(\Lambda = \text{diag}(\lambda_1, \ldots, \lambda_n)\), substituting into the objective gives:
\[
\text{Tr}(U^T L U) = \text{Tr}(Y^T V^T (V \Lambda V^T) V Y) = \text{Tr}(Y^T \Lambda Y),
\]
because \(V^T V = I_n\). Thus:
\[
\text{Tr}(U^T L U) = \sum_{i=1}^n \lambda_i \|Y_{i \cdot}\|_2^2,
\]
where \(Y_{i \cdot}\) denotes the \(i\)-th row of \(Y\). Minimizing \(\text{Tr}(U^T L U)\) means concentrating \(\|Y_{i \cdot}\|_2^2\) onto the smallest \(\lambda_i\). In the extreme, the minimum is achieved when \(U\) spans the eigenvectors corresponding to the \(c\) smallest eigenvalues:
\[
U = [v_1, v_2, \ldots, v_c].
\]

Thus, minimizing \(\text{Tr}(U^T L U)\) forces \(U\) to align with the subspace spanned by the eigenvectors associated with the \(c\) smallest eigenvalues.
Minimizing $\mathrm{Tr}(U^{T}LU)$ aligns $U$ with the eigenvectors associated with the smallest eigenvalues of $L$.

\subsubsection{Connection to Spectral Clustering}

In spectral clustering the Laplacian \(L\) is computed from the similarity graph. then the first \(c\) eigenvectors \(v_1, \ldots, v_c\) corresponding to the smallest eigenvalues are extracted and are used to embed the data points. Clustering (e.g., via \(k\)-means) is performed in this embedded space.

Thus, minimizing \(\text{Tr}(U^T L U)\) during model training directly embeds the clustering behaviour without any need for a separate eigen-decomposition or post-processing clustering step. Spectral clustering uses the smallest eigenvectors of the Laplacian to find clusters. Minimizing \(\text{Tr}(U^T L U)\) embeds this step into the model optimization.

\begin{algorithm}
\caption{Alternating Minimization for the Joint Optimization of $\beta$, $S$, and $U$}
\begin{algorithmic}[1]
\State \textbf{Input:} Data matrices $\{\mathbf{X}_i^k\}_{i=1,\dots,n}^{k=1,\dots,K}$, number of nearest neighbors $\kappa$, parameters $\eta$, $\gamma$, $\lambda$, $\alpha$, $\mu$, and $\{w_k\}_{k=1}^K$.
\State \textbf{Initialize:} Random or heuristic initializations for $\beta$, $S$, and $U$ satisfying constraints.
\While{not converged}
    \State \textbf{Step 1: Update $\beta$ with fixed $S$ and $U$}
    \State Solve the following optimization problem:
    \[
    \min_{\beta} \quad - \sum_{k=1}^{K} \sum_{i=1}^{n} \delta_i^{k} \left( \beta_k^T \mathbf{X}_i^{k} - \log \sum_{j \in \mathcal{R}_i^{k}} \exp\left( \beta_k^T \mathbf{X}_j^{k} \right) \right) + \eta \sum_{k=1}^{K} \|\beta_k\|_1 + \gamma \sum_{i,j} \sum_{k=1}^K w_k \|\beta_k^T (\mathbf{X}_i^k - \mathbf{X}_j^k)\|_2^2 S_{ij}
    \]
    \State Use a proximal gradient method.
    
    \State \textbf{Step 2: Update $S$ with fixed $\beta$ and $U$}
    \For{each $i = 1, \dots, n$}
        \State Solve the following optimization for each row of $S$:
        \[
        \min_{S_{i\cdot}} \quad S_{i\cdot}^T d_i + \lambda \|S_{i\cdot}\|_2^2 \quad \quad s.t. \quad
        \sum_{j=1}^n S_{ij} = 1, \quad S_{ij} \geq 0
        \]
        \State where $d_i$ encodes precomputed weighted distances between sample $i$ and all other samples.
    \EndFor

    \State \textbf{Step 3: Update $U$ with fixed $\beta$ and $S$}
    \State Form the Laplacian matrix $L$ from the current assignment matrix $S$.
    \State Solve the following eigenvalue problem:
    \[
    \min_{U} \quad \mathrm{Tr}(U^T L U) \quad \text{subject to} \quad U^T U = I_c.
    \]
    \State Set $U$ to the eigenvectors corresponding to the $c$ smallest eigenvalues of $L$.
\EndWhile
\State \textbf{Output:} Optimized parameters $\beta$, $S$, and $U$.
\end{algorithmic}
\label{algoritmo}
\end{algorithm}

\subsection{Solution Strategy for the Optimization Problem}

The problem we propose is highly non-convex due to the coupling of \(\beta\), \(S\), and \(U\) within the objective function. However, it is possible to tackle it effectively by using a block coordinate descent approach, also known as \textit{alternating minimization}, where we cyclically optimize over each variable while keeping the others fixed.
Specifically, the method we propose consists of three main steps, repeated iteratively until convergence. A pseudo-code representation is given in Algorithm \ref{algoritmo}. Each step involves optimizing over a single group of variables (\(\beta\), \(S\), or \(U\)) while keeping the others fixed. \\
In our implementation, the transition-specific weights $w_k$ are treated as constant within each iteration~$i$ of the alternating optimization scheme. 
Specifically, at iteration~$i$, the weights are computed using the estimates of the regression coefficients $\beta_k$ obtained at iteration~$i\!-\!1$.
This assumption ensures numerical stability and decouples the estimation of transition weights from the current gradient updates.

\paragraph{Step 1: Update \(\beta\) with \(S\) and \(U\) Fixed}

When the variables \(S\) and \(U\) are fixed, the optimization over \(\beta\) reduces to a regularized optimization problem that combines multiple terms. 
First, there is a survival modelling term based on the log-partial likelihood, which is convex and differentiable in \(\beta\). Then, there is an \(\ell_1\)-norm penalty on \(\beta\), promoting sparsity among the parameters. Finally, there are additional quadratic terms involving \(\beta\) that arise from the similarity-weighted differences between features.
Mathematically, the subproblem in \(\beta\) can be written as:

\[
\begin{aligned}
\min_{\beta}\; \quad
&\underbrace{-\sum_{k=1}^{K}\sum_{i=1}^{n}\delta_{i}^{k}
\left(\beta_{k}^T\mathbf X_{i}^{k}
      -\log \sum_{j\in\mathcal R_{i}^{k}}
      \exp{\left(\beta_{k}^T\mathbf X_{j}^{k}\right)}\right)}_{\text{stratified Cox loss}}\\
&\;+\;
\underbrace{\eta\sum_{k=1}^{K}\lVert\beta_{k}\rVert_{1}}_{\ell_{1}\text{-sparsity}}
\;+\;
\underbrace{\gamma\sum_{k=1}^{K} w_{k}
\sum_{i,j=1}^{n}S_{ij}\left[\beta_{k}^T
      (\mathbf X_{i}^{k}-\mathbf X_{j}^{k})\right]^{2}}_{\text{similarity penalty}}.
\end{aligned}
\]

The first and the last terms are non-linear but differentiable, whereas the \(\ell_{1}\) term is convex but non‑smooth.  
Thus, standard gradient descent cannot be applied directly.
Instead, one effective method is to use proximal methods, such as proximal gradient techniques, where after computing a gradient step on the smooth part of the objective, a proximal mapping is applied to enforce sparsity via soft-thresholding.

Accordingly, we first compute the gradient of the smooth part as:

\[
g(\beta) = \nabla_{\beta_{k}} \left[ -\sum_{k=1}^{K}\sum_{i=1}^{n}\delta_{i}^{k}
\left(\beta_{k}^T\mathbf X_{i}^{k}
      -\log \sum_{j\in\mathcal R_{i}^{k}}
      \exp{\left(\beta_{k}^T\mathbf X_{j}^{k} \right)}\right)
+
\gamma\sum_{k=1}^{K}w_{k}
\sum_{i,j=1}^{n}S_{ij}\left[\beta_{k}^T
      (\mathbf X_{i}^{k}-\mathbf X_{j}^{k})\right]^{2} \right]
\]

that, for every \(k\), becomes:

\begin{align*}
g_k(\beta_k) & = \nabla_{\beta_{k}} \left[
-\sum_{k,i}\delta_{i}^{k}
\left(\beta_{k}^T\mathbf X_{i}^{k}-\log \sum_{j\in\mathcal R_{i}^{k}}
\exp{\left(\beta_{k}^T\mathbf X_{j}^{k} \right)}\right)
+
\gamma\sum_{k}w_{k}\sum_{i,j}S_{ij}
\bigl[\beta_{k}^T(\mathbf X_{i}^{k}-\mathbf X_{j}^{k})\bigr]^{2} \right] \\
 & = -\sum_{i}\delta_{i}^{k}
  \left(\mathbf X_{i}^{k}-\bar{\mathbf X}_{i}^{k}(\beta_{k})\right)
+
2\gamma w_{k}\sum_{i,j}S_{ij} \left[\beta_{k}^T(\mathbf X_{i}^{k}-\mathbf X_{j}^{k})\right]
  \left(\mathbf X_{i}^{k}-\mathbf X_{j}^{k} \right).
\end{align*}

Intuitively the first part is the difference between the covariate of the
individual who failed and the risk‑set average at the same time point,
and drives the fit toward covariates that increase the partial likelihood.
The second term measures how far apart patients
\(i\) and \(j\) are in the current \(\beta_{k}\)-induced risk space,
while the outer factor \((\mathbf X_{i}^{k}-\mathbf X_{j}^{k})\) nudges
\(\beta_{k}\) so as to shrink that distance whenever \(S_{ij}\) says the pair should be considered similar.

Being \(\alpha_{k}\) a fixed step size, the update reads
\[
\beta_{k}^{(t+1)}=
\operatorname{prox}_{\alpha_{k}h}\!
\Bigl(
\beta_{k}^{(t)}-\alpha_{k}\,g_{k}\bigl(\beta_{k}^{(t)}\bigr)
\Bigr).
\]
Since \(h\) is the weighted \(\ell_{1}\) norm, its proximal operator is
the element‑wise \emph{soft threshold},
\[
\bigl[\operatorname{prox}_{\alpha_{k}\eta\lVert\cdot\rVert_{1}}(v)\bigr]_{p}
=
\operatorname{sign}(v_{p})
\,\max\!\bigl(|v_{p}|-\alpha_{k}\eta,\;0\bigr).
\]
Geometrically, every coordinate of
\(v=\beta_{k}^{(t)}-\alpha_{k}g_{k}(\beta_{k}^{(t)})\) is
pulled toward zero by the amount \(\alpha_{k}\eta\); if the pull is
stronger than the coordinate’s magnitude the coefficient is set
\emph{exactly} to zero, realising sparsity.

\paragraph{Step 2: Update \(S\) with \(\beta\) and \(U\) Fixed}

When \(\beta\) and \(U\) are held fixed, the optimization over \(S\) simplifies significantly into the following subproblem:

\begin{equation}
\label{eq:min_over_S}
\begin{aligned}
    &\min_{S} 
    \underbrace{\quad \gamma \sum_{i,j} \left( \mu \|\mathbf{X}_i - \mathbf{X}_j\|_2^2 + \sum_{k=1}^K w_k \|\beta_k^T (\mathbf{X}_i^k - \mathbf{X}_j^k)\|_2^2 \right) S_{ij}}_{\text{linear term}} + 
    \underbrace{\lambda \sum_{i,j} S_{ij}^2}_{\text{quadratic term}}\\
    &s.t.
    \underbrace{\sum_{j} S_{ij} = 1}_{\text{equality constraint}}, \quad \underbrace{S_{ij} \geq 0}_{\text{inequality constraint}}.
\end{aligned}
\end{equation}

The terms depending on \(S\) are linear and quadratic, while the constraints on \(S\) are simply an equality and an inequality linear constraints respectively. 
As the optimization for each row \(S_{ij}\) is independent of the others and can be addressed independently, the optimization of such problem decomposes across the rows of \(S\): for each \(i\), we are solving a simple constrained quadratic minimization problem in \(n\) variables. 
This objective function is thus convex in \(S_{ij}\) and, by the fundamental theorem of convex optimization, 
such convexity and the constraints imply that the optimal solution is unique, i.e., there is a single point where the gradient is zero and the constraints are satisfied.

We streamline the formulation of the objective function by defining:

\[
d_{ij} := \mu \|\mathbf{X}_i - \mathbf{X}_j\|_2^2 + \sum_{k=1}^K w_k \|\beta_k^\top (\mathbf{X}_i^k - \mathbf{X}_j^k)\|_2^2.
\]

so that Equation \ref{eq:min_over_S} becomes:

\[
\min_{S} \quad \gamma \sum_{ij} d_{ij} S_{ij} + \lambda \sum_{ij} S_{ij}^2,
\quad \text{subject to} \quad \sum_j S_{ij} = 1, \quad S_{ij} \geq 0,
\]

To find the solution, we convert the constrained problem into an unconstrained problem where to apply the derivative test using a Lagrangian expression \cite{guo2015robust}. 
First, we introduce a Lagrange multiplier \( \eta \in \mathbb{R} \) to handle the equality constraint:

\[
\mathcal{L}(S, \eta) =  \gamma \sum_{ij} d_{ij} S_{ij} + \lambda \sum_{ij} S_{ij}^2 - \eta \left( \sum_{ij} S_{ij} - \mathbf{1} \right),
\]


where \( \mathbf{1} \in \mathbb{R}^n \) is the vector of ones, which can be simplified into:

\[
\mathcal{L}(S, \eta) =  \sum_{ij} \left( \gamma d_{ij} S_{ij} + \lambda S_{ij}^2 - \eta S_{ij} \right) + \eta \cdot \mathbf{1}
\]

To find the solution \(S_{ij}\), we take the derivative of the Lagrangian with respect to \(S_{ij}\) and set it equal to zero:

\[
\frac{\partial \mathcal{L}}{\partial S_{ij}} = \gamma d_{ij} + 2\lambda S_{ij} - \eta = 0.
\]

Thus, from the first-order optimality conditions, the derive the closed-form solution \(S_{ij}\) by solving the above equation in terms of the parameters of the problem:
\[
S_{ij} = \frac{\eta - \gamma d_{ij}}{2\lambda}.
\]

Finally, to comply with the non-negativity constraint \(S_{ij} \geq 0\), which would require further adjustments via a projection step (i.e., soft-thresholding or clamping negative values to zero), we use the notation $(a)_+ = \max(a, 0)$ to identify the function that converts the values of \(a\) smaller than $0$ to $0$. Specifically, we project the solution onto the positive subspace:

\begin{equation}
    \label{eq:solution_S}
    S_{ij} = \left( \frac{\eta - \gamma d_{ij}}{2\lambda} \right)_+,
\end{equation}

where \( (a)_+ = \max(a, 0) \) is applied elementwise.

\paragraph{Step 3: Update \(U\) with \(\beta\) and \(S\) Fixed}

Finally, when \(\beta\) and \(S\) are fixed, the optimization over \(U\) involves only the trace term:

\[
\min_{U} \quad \mathrm{Tr}(U^{T} L U) \quad \text{subject to} \quad U^T U = I_c,
\]
where \(L\) is the Laplacian matrix computed from the current \(S\).
This is a classical spectral optimization problem. The Laplacian \(L\) is a symmetric positive semi-definite matrix, and the objective function seeks an orthonormal matrix \(U\) that minimizes the quadratic form associated with \(L\).

It is well-known that the solution is given by taking the eigenvectors of \(L\) associated with its \(c\) smallest eigenvalues. That is, we perform an eigen-decomposition of \(L\), and we form the matrix \(U\) by stacking the first \(c\) eigenvectors column-wise. In practice, we do not need to compute the full eigen-decomposition, but rather only the first few eigenvectors, which can be done efficiently using methods like iterative eigensolvers \cite{nie2014clustering}.




\paragraph{Adaptive neighbor parameter tuning}

In the context of solving the optimization problem in Equation \ref{eq:min_over_S} according to Equation \ref{eq:solution_S}, it is beneficial to promote sparsity in the learned similarity matrix \(S\) by encouraging each data point \(i\) to connect only to a small number of neighbors. 
Indeed, rather than assigning nonzero similarity weights to all pairs \((i,j)\), a sparse assignment focusing only on the $\kappa$-nearest neighbors improves robustness and better reflects the local geometry of the data manifold while reducing the computational complexity. This can be achieved by carefully tuning  the regularization parameter \(\lambda\) (and \(\eta\)) for each data point \(i\).

To enforce that each point \( i \) connects only to its $\kappa$-nearest neighbors, we impose that \( S_{ij} > 0 \) if and only if \( j \in J_{\kappa} \), where \( J_{\kappa} \) denotes the index set of the $\kappa$-nearest neighbors of point \( i \).
This requirement translates into the condition:

\begin{equation}
\label{eq:k-ineq}
S_{ij} = \left\{
\begin{array}{ll}
\frac{\eta - \gamma d_{ij}}{2\lambda} > 0, & \text{if } j \in J_{\kappa}, \\
0, & \text{if } j \notin J_{\kappa}.
\end{array}
\right.
\end{equation}

From the positivity condition for \( j \in J_{\kappa} \), and using the constraint \( \sum_{j \in J_{\kappa}} S_{ij} = 1 \), we derive:

\begin{align}
\sum_{j \in J_{\kappa}} S_{ij} &= \sum_{j \in J_{\kappa}} \frac{\eta - \gamma d_{ij}}{2\lambda} \\
&= \frac{1}{2\lambda} \left( k\eta - \gamma \sum_{j \in J_{\kappa}} d_{ij} \right) = 1.
\end{align}

Solving for \( \eta \), we obtain:

\begin{equation}
\eta = \frac{1}{\kappa} \left( 2\lambda + \gamma \sum_{j \in J_{\kappa}} d_{ij} \right).
\end{equation}

To ensure that \( S_{ij} = 0 \) for \( j \notin J_{\kappa} \), the argument inside the positive part of \eqref{eq:solution_S} must be non-positive:

\begin{align}
0 &= \left( \frac{\eta - \gamma d_{ij}}{2\lambda} \right), \quad j \notin J_{\kappa}, \\
\Rightarrow \eta &= \gamma d_{ij}, \quad j \notin J_{\kappa}.
\end{align}

Combining this with the previous expression for \( \eta \), we derive:

\begin{equation}
\gamma d_{ij} = \frac{1}{\kappa} \left( 2\lambda + \gamma \sum_{j' \in J_{\kappa}} d_{ij'} \right), \quad j \notin J_{\kappa},
\end{equation}

from which we solve for \( \lambda \):

\begin{align}
2\lambda &= \gamma \kappa d_{ij} - \gamma \sum_{j' \in J_{\kappa}} d_{ij'} \\
\Rightarrow \lambda &= \frac{\gamma}{2} \left( \kappa d_{ij} - \sum_{j' \in J_{\kappa}} d_{ij'} \right).
\end{align}

The final solution would be:

\begin{equation}
S_{ij} = \left\{
\begin{array}{ll}
\displaystyle \frac{1}{2\lambda} \left( \frac{1}{\kappa} \left( 2\lambda + \gamma \sum_{j' \in J_{\kappa}} d_{ij'} \right) - \gamma d_{ij} \right), & j \in J_{\kappa}, \\
0, & j \notin J_{\kappa},
\end{array}
\right.
\end{equation}

A key advantage of this formulation is that it removes the need to specify the number of clusters \(c\) a priori, as typically required in clustering-based models. Instead, the graph structure encoded by \(S\) naturally emerges from the local relationships in the data without enforcing a global partitioning beforehand. This leads to a more flexible and data-driven learning process.

\subsection{Computational complexity}

The computational complexity of one iteration of the external while loop is given by the sum of the complexities of the estimation of \(\beta\) and \(S\).
The complexity for the estimation of \(\beta\) is the one of the proximal gradiend descent algorithm: \(\mathcal{O}(i\ n \sum_k p^k)\), where \(p^k\) is the number of features of transition \(k\), \(n\) the number of patients, \(i\) the number of iterations performed in the inner \emph{while} loop. 
The complexity for the estimation of the similarity matrix \(S\) is \(\mathcal{O}(n^2)\).
Named \(l\) the number of iterations of the outer loop, the overall complexity is \(\mathcal{O}(l(i\ n \sum p^k + n^2))\).

\section{Simulation study}
\label{sec:simulation-study}

To evaluate the performance and robustness of the proposed model under a range of realistic and challenging settings, we designed a comprehensive simulation study that systematically varies key data characteristics and methodological parameters. 
In the following sections, we first introduce how we generate the data depending on experimental parameters (Section \ref{subsec:data-generation}), then we detail how we vary such parameters to perform different tests on our model and principal benchmark models from literature according to major existing performance evaluation metrics (Section \ref{subsec:experimental-setup}), and finally we conduct an ablation study as inquire wheter  our model retains all and only the useful terms (Section \ref{subsec:ablation-study}).

\subsection{Data generation}
\label{subsec:data-generation}

We generate synthetic datasets according to a multi-step pipeline designed to simulate clustered, high-dimensional, time-to-event data with multi-state progression. 
First, latent patient clusters are assigned by drawing cluster sizes from a multinomial distribution: 
\[ 
(n_1, \ldots, n_C) \sim \mathrm{Multinomial}(n, \pi),
\]

where \( \pi_c = 1/C \) by default, ensuring balanced cluster sizes. To investigate the impact of class imbalance, this prior can be adjusted. Each patient \( i \in \{1, \ldots, n\} \) is then assigned to a cluster \( c(i) \in \{1, \ldots, C\} \).

For each patient, covariate vectors \( \mathbf{X}_i \in \mathbb{R}^p \) are drawn conditional on cluster membership. Specifically, within each cluster \( c \), covariates follow a multivariate normal distribution: 
\[  
\mathbf{X}_i \sim \mathcal{N}(\mu_c, \Sigma),
\]

where \( \Sigma \) is initially taken to be diagonal with entries \( \sigma_1^2, \ldots, \sigma_p^2 \), but may optionally include a correlation structure, e.g., \( \Sigma_{jk} = \rho^{|j-k|} \), allowing for dependence across features. 
The cluster-specific mean vectors \( \mu_c \) are drawn once at the beginning of each repetition with elements \( \mu_{cj} \sim \mathcal{N}(0, \tau^2) \), where the variance parameter \( \tau^2 \) controls the degree of separation between clusters in the covariate space.

To simulate event times across the sequence of states (i.e., 0 \( \rightarrow \) 1 \( \rightarrow \) 2), we specify a set of transition-specific hazard functions governed by log-linear covariate effects. For each transition \( k \), we generate the true effect vector \( \beta_k^* \in \mathbb{R}^p \) as a combination of a shared effect and a sparse, transition-specific deviation: 
\[ 
\beta_k^* = \beta_{\text{shared}} + \delta_k,
\]
where \( \beta_{\text{shared}} \sim \mathcal{N}(0, I_p) \) encodes global risk factors, and \( \delta_k \) has \( s\% \) of its coordinates drawn from \( \mathcal{N}(0,1) \), with the remainder set to zero, promoting sparsity and ensuring that the $\ell_1$-penalty in model fitting is meaningful. 

Event times are drawn from transition-specific Weibull distributions. The baseline hazard for transition \( k \) is given by 
\[
h_{0k}(t) = \alpha_k \nu_k t^{\nu_k - 1}.
\]

For patient \( i \), the individual-specific hazard is computed as 
\[ 
h_{ik}(t) = h_{0k}(t) \exp(\beta_k^{*T} X_i).
\]

Event times are sampled using inverse transform sampling: 
\[
T_{ik} = \left( -\frac{\log U}{\alpha_k \exp(\beta_k^{*T} X_i)} \right)^{1/\nu_k}, \quad U \sim \mathcal{U}(0,1),
\]
and sequential progression through states is enforced by summing successive transitions, e.g., \( T_{i,k+1} := T_{ik} + T'_{i,k+1} \), ensuring temporal consistency in the multi-state process. To simulate right-censoring, censoring times \( C_i \sim \mathcal{U}(0, q) \) are drawn, with the parameter \( q \) calibrated to yield a desired average censoring rate (e.g., 25\%). 

The observed survival times and event indicators are then defined as:
\begin{equation}
\left\{ \begin{array}{l}
Y_{ik} = \min(T_{ik}, C_i) \\
\delta_{ik} = \mathbb{I}[T_{ik} \leq C_i].
\end{array}\right.
\end{equation}

Finally, we compute a ground-truth similarity matrix \( S^* \in \mathbb{R}^{n \times n} \) that reflects both covariate and risk profile proximity, for evaluating edge-level recovery and clustering performance. 
This matrix reflects both covariate proximity and risk profile similarity and is defined as:
\[
S_{ij}^* = \frac{\exp\left( -\sigma_X \| X_i - X_j \|_2^2 - \sigma_\lambda \sum_k \left| \beta_k^{*T} (X_i - X_j) \right| \right)}{\sum_{j'} \exp\left( -\sigma_X \| X_i - X_{j'} \|_2^2 - \sigma_\lambda \sum_k \left| \beta_k^{*T} (X_i - X_{j'}) \right| \right)}.
\]

Specifically, $S^*$ is computed via a softmax kernel over pairwise distances in covariate space and projected risk space, parameterized by hyperparameters \( \sigma_X \) and \( \sigma_\lambda \). This approach provides a continuous, structure-aware benchmark for graph recovery, inspired by affinity-based graph learning methods \cite{zhou2004learning, belkin2003laplacian, wang2009manifold, wang2020graph}. Unlike binary cluster labels, this allows for fine-grained evaluation of the learned graph’s fidelity to the true underlying structure generated in our simulation framework.
The hyperparameters \( \sigma_X \) and \( \sigma_\lambda \) are tuned so that, on average, within-cluster similarities \( S_{ij}^* \) are approximately 0.05 higher than between-cluster similarities, providing a useful external benchmark for graph recovery. This matrix enables computation of metrics such as edge prediction AUC and clustering agreement (e.g., ARI, AMI) without relying on the graph learned by the model under evaluation.

\subsection{Experiments}
\label{subsec:experimental-setup}


We designed targeted simulation experiments to validate key properties of the proposed model. Each benchmark focuses on a specific hypothesis, uses a tailored evaluation metric, and tests whether the observed performance aligns with theoretical expectations. This section summarizes our methodology and findings across the main dimensions of model evaluation.

\subsubsection{Set up}

Table \ref{tab:simulation-summary} summarizes the dimensions explored, including the number of patients ($n$), number of covariates ($p$), number of latent disease states ($K$), number of underlying patient clusters ($C$), baseline hazard types, and censoring rates. 

\begin{table}[h]
    \centering
    \begin{tabularx}{\linewidth}{XllX}
        \toprule
        \textbf{Dimension} & \textbf{Default} & \textbf{Values to sweep} & \textbf{Rationale} \\
        \hline
        Patients ($n$) & 500 & 200 -- 2\,000 & effects of sample size \& sparsity \\
        Covariates ($p$) & 30 & 10, 50, 100 & high-dimensional regime \\
        States ($K$) & 3 (0 $\rightarrow$ 1 $\rightarrow$ 2) & 2, 4 & typical chronic-disease paths \\
        Clusters ($C$) & 4 & 2, 6 & hidden strata you hope to uncover \\
        Censoring rate & 25\% & 0\%, 40\% & realistic right-censoring \\
        Replicates & 30 & -- & stable confidence intervals \\
        \bottomrule
    \end{tabularx}
    \caption{Simulation dimensions, defaults, and rationale for parameter sweeps.}
    \label{tab:simulation-summary}
\end{table}

Each experiment follows a structured workflow to ensure fair comparison and reproducibility. For every combination of dataset parameters 
we repeat the experiment 30 times to account for stochastic variability.
In each repetition, we simulate a synthetic dataset comprising covariates $X$, event times $Y$, censoring indicators $\delta$, true underlying clusters, and ground truth state trajectories $S^*$. Following our modelling approach, we conduct a grid search to tune the hyperparameters $(\gamma, \mu, \lambda, k)$, and the optimal values are then used to train the model. Performance is evaluated using a comprehensive set of metrics.
Table~\ref{tab:evaluation-metrics} summarizes the evaluation metrics used to assess various aspects of model performance, including predictive accuracy, graph structure recovery, clustering quality, and computational efficiency.
For benchmarking, we compare against the baseline and reference models listed in Table~\ref{tab:baselines}. These models are trained and evaluated under the same experimental settings as our proposed method, using identical input dimensions and evaluation criteria.
As each experiment is repeated 30 times, we report the results as summary statistics (mean $\pm$ standard deviation) of the evaluation metrics. 

\begin{table}[h]
    \centering
    \begin{tabularx}{\linewidth}{XX}
        \toprule
        \textbf{Metric} & \textbf{Description} \\
        \hline
        \multicolumn{2}{c}{\textit{Survival-prediction quality}} \\
        \hline
        Concordance Index (C-index) & Measures the concordance between predicted and observed event times for each transition. Higher values indicate better discrimination. \\ \addlinespace
        Time-dependent AUROC & Optional metric capturing the model’s discriminatory power at different time points. \\
        \hline
        \multicolumn{2}{c}{\textit{Graph / clustering recovery}} \\
        \hline
        Adjusted Rand Index (ARI) & Measures the agreement between cluster assignments obtained from spectral clustering on \( S \) and true latent clusters \( c(i) \). \\ \addlinespace
        Adjusted Mutual Information (AMI) & Quantify clustering agreement with a focus on completeness and homogeneity. \\
        \hline
        \multicolumn{2}{c}{\textit{Model characteristics}} \\
        \hline
        Sparsity ratio \( \|\beta\|_0 / p \) & Evaluates the fraction of active features; used to assess the impact of the \( \ell_1 \)-penalty. \\ \addlinespace
        Runtime and memory scaling & Reports wall-clock time and memory footprint, especially focusing on \( \mathcal{O}(n^2) \) operations (e.g., on \( n = 2000 \)). \\
        \bottomrule
    \end{tabularx}
    \caption{Evaluation metrics used to assess survival prediction performance, graph/clustering recovery, and computational characteristics.}
    \label{tab:evaluation-metrics} 
\end{table}

\begin{table}[h]
    \centering
    \begin{tabularx}{\linewidth}{XX}
        \toprule
        \textbf{Label} & \textbf{Description} \\
        \hline 
        \textbf{Our model} & Full joint optimization model that learns both the regression coefficients \( \beta \) and the similarity matrix \( S \) simultaneously. \\ \addlinespace
        \textbf{Cox-only} & Stratified Cox model using the same loss structure but without the similarity regularization term (\( \gamma = 0 \)). \\ \addlinespace
        \textbf{Fixed-Graph + Cox} & Uses a pre-computed RBF similarity graph based on covariates; optimizes only the Cox model coefficients \( \beta \). \\ \addlinespace
        \textbf{k-NN + Cox} & Similar to Fixed-Graph + Cox but employs a sparse binary graph constructed via $k$-nearest neighbors. \\ \addlinespace
        \textbf{Random Survival Forest} & A strong non-parametric ensemble baseline that handles censoring without explicit model structure. \\ 
        \bottomrule
    \end{tabularx}
    \caption{Baseline and comparative models used in the experimental study.}
    \label{tab:baselines}
\end{table}

\subsubsection{Experiment 1: Joint learning improves prediction}

We first evaluate the robustness of the survival prediction depending on the simulation dimensions we defined in Table \ref{tab:simulation-summary}.
Moreover, we assess whether jointly optimizing the covariate effects \( \beta \) and the similarity graph \( S \) yields better predictive performance than comparative methods. We use the concordance index (C-index) and time-dependent AUROC. 
Results are displayed in Table \ref{tab:results-exp1}.

Jointly estimating the covariate effects and the similarity matrix enhances survival prediction compared to all competing approaches. The proposed model consistently achieves the highest C-index and time-dependent AUROC, clearly outperforming the comparison baselines. This confirms the benefit of coupling representation learning of patient similarity with risk modelling rather than relying on a pre-defined graph. Robustness analysis across simulation dimensions (number of patients, covariates, latent states, and censoring rates) shows that performance remains stable, with only minor fluctuations, attesting to the model’s resilience to structural noise in the data.

\begin{table}[h!]
\centering
\begin{tabularx}{\textwidth}{lXXX}
\toprule
\textbf{} & \textbf{C-index} & \textbf{Time dependent AUROC} \\
\midrule
\multicolumn{3}{c}{\textit{Robustness}} \\
\hline
Patients (n)            &0.939\(\pm\)0.011&0.984 \(\pm\) 0.013       \\
Covariates (p)          &0.925 \(\pm\) 0.019&0.982 \(\pm\) 0.011       \\
States (K)              &0.937 \(\pm\) 0.011&0.970 \(\pm\) 0.023       \\
Clusters (C)            &0.942 \(\pm\) 0.013&0.985 \(\pm\) 0.011   \\
Censoring rate (q)      &0.945 \(\pm\) 0.017&0.986 \(\pm\) 0.01    \\
Variance parameter ($\tau$)  &0.924 \(\pm\) 0.015&0.964 \(\pm\) 0.015       \\
\addlinespace
\hline
\multicolumn{3}{c}{\textit{Comparison (with default values)}} \\
\hline
Our model                &0.977 \(\pm\) 0.007&0.988 \(\pm\) 0.001\\
Cox-only                 &0.548 \(\pm\) 0.007&0.735 \(\pm\)0.038\\
Fixed-Graph + Cox        &0.507 \(\pm\) 0.09&0.712 \(\pm\) 0.036\\
k-NN + Cox               &0.582 \(\pm\) 0.009&0.754 \(\pm\)0.033\\
Random Survival Forest   &0.81 \(\pm\) 0.04&0.887 \(\pm\) 0.007\\
\bottomrule
\end{tabularx}
\caption{Experiment 1: Robustness and Model Comparison; results are presented in terms of Confidence Intervals of the metrics, as obtained from 30 replicas of the experiments.}
\label{tab:results-exp1}
\end{table}

\subsubsection{Experiment 2: Graph term recovers latent strata}

To evaluate the ability of the learned graph to capture underlying patient heterogeneity, we compare cluster assignments obtained via spectral clustering on the similarity matrix \( S \) to the true latent groupings. The adjusted Rand index and the Adjusted Mutual Information (AMI) serve as the metric of interest. We test the model on data with varying parameters (with reference to Table \ref{tab:evaluation-metrics}) and we compared it to competing models, discussing successful recovery of latent strata.
Results are presented in Table \ref{tab:results-exp2}.

The proposed model consistently attains equal-to-higher Adjusted Rand Index (ARI) and Adjusted Mutual Information (AMI) than competing graph-based or tree-based baselines, confirming that the learned similarity matrix S captures the true underlying patient structure. When the separation among clusters decreases (higher \(\tau\)), both ARI and AMI decline, as expected, since the latent strata become less distinct. Overall, the results are suggesting that the learned graph embeds both covariate and risk-profile similarities effectively.
\begin{table}[h!]
\centering
\begin{tabularx}{\textwidth}{lXXX}
\toprule
\textbf{} & \textbf{ARI} & \textbf{AMI} \\
\midrule
\multicolumn{3}{c}{\textit{Robustness}} \\
\hline
Patients (n)            &0.02 \(\pm\) 0.013&0.02 \(\pm\) 0.013       \\
Covariates (p)          &-0.003 \(\pm\) 0&0.018 \(\pm\) 0.009       \\
States (K)              &0.002 \(\pm\) 0.006&0.017 \(\pm\) 0.009\\
Clusters (C)            &0.013 \(\pm\) 0.015&0.013 \(\pm\) 0.007       \\
Censoring rate (q)      &0.007 \(\pm\) 0&0.012 \(\pm\) 0.006       \\
Variance parameter ($\tau$)  &0 \(\pm\) 0&0.018 \(\pm\) 0.008       \\
\addlinespace
\hline
\multicolumn{3}{c}{\textit{Comparison (with default values)}} \\
\hline
Our model                &0.015 \(\pm\) 0.030&0.007 \(\pm\) 0.010\\
Cox-only                 &N.A.       &N.A.       \\
Fixed-Graph + Cox        & 0 \(\pm\) 0.01 & 0.023 \(\pm\) 0.002 \\
k-NN + Cox               &0.002 \(\pm\) 0.006  & 0.004 \(\pm\) 0.004   \\
Random Survival Forest   &N.A.&N.A.\\
\bottomrule
\end{tabularx}
\caption{Experiment 2: Robustness and Model Comparison; results are presented in terms of Confidence Intervals of the metrics, as obtained from 30 replicas of the experiments.}
\label{tab:results-exp2}
\end{table}

\subsubsection{Experiment 3: $\ell_1$ finds right features and allows scaling}

We test whether the sparsity-inducing penalty correctly identifies the subset of covariates with non-zero effects. This is quantified via the estimated sparsity ratio \( \|\beta\|_0/p \), which we compare to the true signal proportion $s$ (see Table \ref{tab:evaluation-metrics}). 
Moreover, to validate the scalability of our method, we measure runtime and memory scaling as a function of sample size. 
Results are examined across varying dimensions and are presented in Table \ref{tab:results-exp3}.

The runtime analysis exhibits a near-linear trend in log-log scale with respect to sample size. Memory usage remains stable, confirming that the algorithm scales efficiently to larger cohorts without excessive computational overhead.

\begin{figure}
    \centering
    \includegraphics[width=0.7\linewidth]{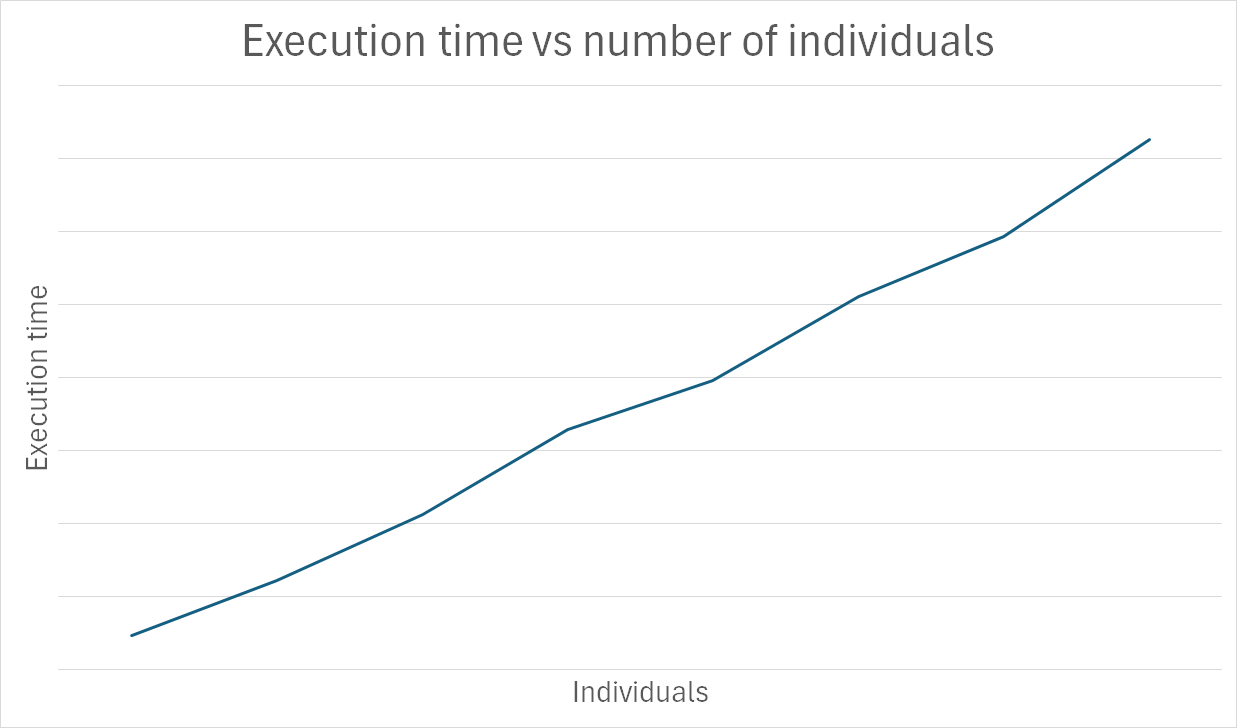}
    \caption{Log-log plot of the execution time of the proposed algorithm as a function of the number of individuals in the study, ranging from 100 to 800. Data points represent average execution times measured over 30 runs. The linear trend on the log-log scale suggests scalability characteristics of the algorithm across varying population sizes.}
    \label{fig:execution-time}
\end{figure}


\begin{table}[h!]
\centering
\begin{tabularx}{\textwidth}{lXXX}
\toprule
\textbf{} & Sparsity ratio & Runtime & Memory \\
\midrule
\multicolumn{4}{c}{\textit{Robustness}} \\
\hline
Patients (n)            &1 \(\pm\)0&11.94 \(\pm\) 8.04&138 \(\pm\) 3       \\
Covariates (p)          &1 \(\pm\) 0&28.13 \(\pm\) 19.52&134 \(\pm\) 1       \\
States (K)              &1 \(\pm\) 0&4.95 \(\pm\) 1.24&134 \(\pm\) 1\\
Clusters (C)            &1 \(\pm\) 0&8.25 \(\pm\) 2.92&134 \(\pm\) 1       \\
Censoring rate (q)      &1 \(\pm\) 0&9.70 \(\pm\) 3.83&134 \(\pm\) 0       \\
Variance parameter ($\tau$)  & 1 \(\pm\) 0&9.39 \(\pm\) 3.63&134 \(\pm\)1       \\
\addlinespace
\hline
\multicolumn{4}{c}{\textit{Comparison (with default values)}} \\
\hline
Our model                &1 \(\pm\) 0&3.71 \(\pm\) 0.01&129 \(\pm\) 1\\
Cox-only                 &1 \(\pm\) 0&0.72 \(\pm\)0.01&130 \(\pm\) 1\\
Fixed-Graph + Cox        & 1 \(\pm\) 0& 0.39 \(\pm\) 0.02 & 134 \(\pm\) 0\\
k-NN + Cox               &1 \(\pm\) 0  &0.42 \(\pm\) 0.01&133 \(\pm\) 1\\
Random Survival Forest   &1 \(\pm\) 0&0.22 \(\pm\) 0.017&150 \(\pm\) 1\\
\bottomrule
\end{tabularx}
\caption{Experiment 3: Robustness and Model Comparison; results are presented in terms of Confidence Intervals of the metrics, as obtained from 30 replicas of the experiments.}
\label{tab:results-exp3}
\end{table}

\subsection{Ablation study}
\label{subsec:ablation-study}

To better understand the contribution of each component of our model, we conducted a series of ablation studies. In each setting, we modify a single aspect of the modelling or data-generating pipeline while keeping all other factors constant. 
To assess the importance of the different regularization terms, we individually disable each component: the similarity-based graph penalty (\( \gamma = 0 \)), the neighbor smoothness regularizer (\( \lambda = 0 \)), and the sparsity-promoting \( \ell_1 \)-penalty (\( \eta = 0 \)). This helps isolate the contribution of each inductive bias to overall model performance, particularly in terms of survival discrimination and latent cluster recovery. We report the resulting change in performance with respect to the full model using two key metrics: C-index (for survival prediction) and ARI (for clustering recovery).
Results are presented in Table \ref{tag:ablation-results}.

Disabling any of the components results in performance degradation, substantially lowering both C-index and clustering quality, emphasizing their complementary roles in capturing population structure and ensuring consistent survival estimation. The full model therefore represents a balanced combination of components, each contributing to stability and generalization.

\begin{table}[h!]
\centering
\begin{tabularx}{\textwidth}{lXX}
\toprule
 & \textbf{C-index} & \textbf{ARI} \\
\midrule
$\gamma = 0$&0.950 \(\pm\) 0.017&0.017 \(\pm\) 0.007\\
$\lambda = 0$ & 0.950 \(\pm\) 0.014&0.000 \(\pm\)0.001\\
$\eta = 0$&0.970 \(\pm\) 0.006&-0.110 \(\pm\)0.010\\
Full model&0.977 \(\pm\) 0.007&0.015 \(\pm\) 0.030\\
\bottomrule
\end{tabularx}
\caption{Ablation study: Effect of Hyperparameters on Model Performance}
\label{tag:ablation-results}
\end{table}

\section{Application to Liver Metastases (LM)}
\label{sec:application}

We further validate our method with a real world case study, in the context of Liver Metastases (LM).
Metastatic progression is a leading cause of cancer-related mortality, with the liver representing the most common site of distant metastases. In colorectal cancer (CRC), approximately 50\% of patients develop colorectal liver metastases (CRLM) during the course of the disease. While systemic chemotherapy and surgical resection remain the primary therapeutic options, surgical eligibility is limited to a small subset of patients, and treatment selection is largely empirical \cite{zhou2022colorectal, chow2019colorectal}. Currently, no validated biomarkers exist to stratify CRLM into clinically meaningful subtypes that could inform therapy. Given the large use of contrast-enhanced computed tomography (CT) in routine oncology workflows, imaging data represent a rich, underutilized source of non-invasive phenotypic information with potential biomarker utility \cite{fiz2020radiomics, baghdadi2022imaging}. In this case study, we utilize time-resolved imaging and longitudinal clinical data to evaluate a trajectory-informed cancer subtyping framework. Our aim is to characterize imaging-based features of disease evolution that may serve as candidate biomarkers, enabling future work in predictive modelling and personalized therapy selection.

\subsection{Dataset description}

This study is based on a retrospective cohort of colorectal cancer patients treated at the Humanitas Research Hospital in Rozzano, Italy. The dataset comprises 102 patients diagnosed with colorectal liver metastases (CRLM), each having undergone systemic chemotherapy followed by hepatic resection.
The collection and analysis of patient data were carried out under approval from the Institutional Review Board of Humanitas Research Hospital
, in compliance with the Declaration of Helsinki and relevant national regulations on biomedical research. Because of the retrospective nature of the study, the need for informed consent was waived.

For each patient, baseline clinical variables were recorded prior to the beginning of chemotherapy. These include age (in years), sex (binary: male/female), number of liver metastases (discrete count), presence of extrahepatic disease (binary: 0 = absent, 1 = present), timing of liver metastasis (binary: synchronous vs. metachronous with respect to primary tumour diagnosis).
These covariates are commonly used in clinical practice to stratify risk and estimate prognosis and are therefore included for both descriptive and modelling purposes. 
For each patient, the largest liver metastasis was identified and analyzed using contrast-enhanced computed tomography (CT) scans at two time points: Baseline ($T_0$), prior to the initiation of systemic chemotherapy; Post-treatment ($T_1$), following completion of chemotherapy but prior to surgical resection.
In both CT acquisitions, the lesion was manually segmented by expert radiologists according to standardized oncologic imaging protocols using LIFEx sotware (\cite{nioche2018lifex},\href{https://www.lifexsoft.org/}{www.lifexsoft.org}). From each segmented lesion, 109 radiomic features were extracted using a validated pipeline, which includes intensity, texture, and shape descriptors consistent with the Image Biomarker Standardization Initiative (IBSI) guidelines \cite{zwanenburg2016image}. These features are designed to quantitatively capture intra-tumoural heterogeneity and lesion morphology and are denoted by the vector:
\[
\mathbf{X}^{(t)}_i = \left(X^{(t)}_{i,1}, X^{(t)}_{i,2}, \ldots, X^{(t)}_{i,109}\right) \in \mathbb{R}^{109}, \quad t \in \{0,1\}, \quad i \in \{1,...,n\},
\]
where $\mathbf{X}^{(t)}_i$ represents the radiomic feature vector of the $i$-th patient at time $t$ (baseline or post-treatment).

All patients underwent surgical resection following chemotherapy. We collected information about two main clinical outcomes, i.e., \textit{relapse}, defined as the first occurrence of recurrent disease following resection, and \textit{overall survival}, defined as time to death from any cause following resection.
For each patient, the occurrence (event indicator) and timing (in months) of relapse and death were recorded, yielding right-censored survival outcomes. Let $T^{R}_i$ and $T^{D}_i$ denote the time-to-relapse and time-to-death, respectively, with corresponding censoring indicators $\delta^{R}_i, \delta^{D}_i \in \{0,1\}$.

Additionally, we recorded information about treatment-related variables including the length of chemotherapy (in months) and the time interval between end of chemotherapy and resection (in weeks). These temporal markers provide context for interpreting both radiomic changes and outcome events.
A complete summary of the clinical variables, treatment durations, radiomic feature distributions, and event time statistics is provided in the Supplementary Materials.

\begin{figure}[h!]
    \centering
    \includegraphics[width=0.85\textwidth]{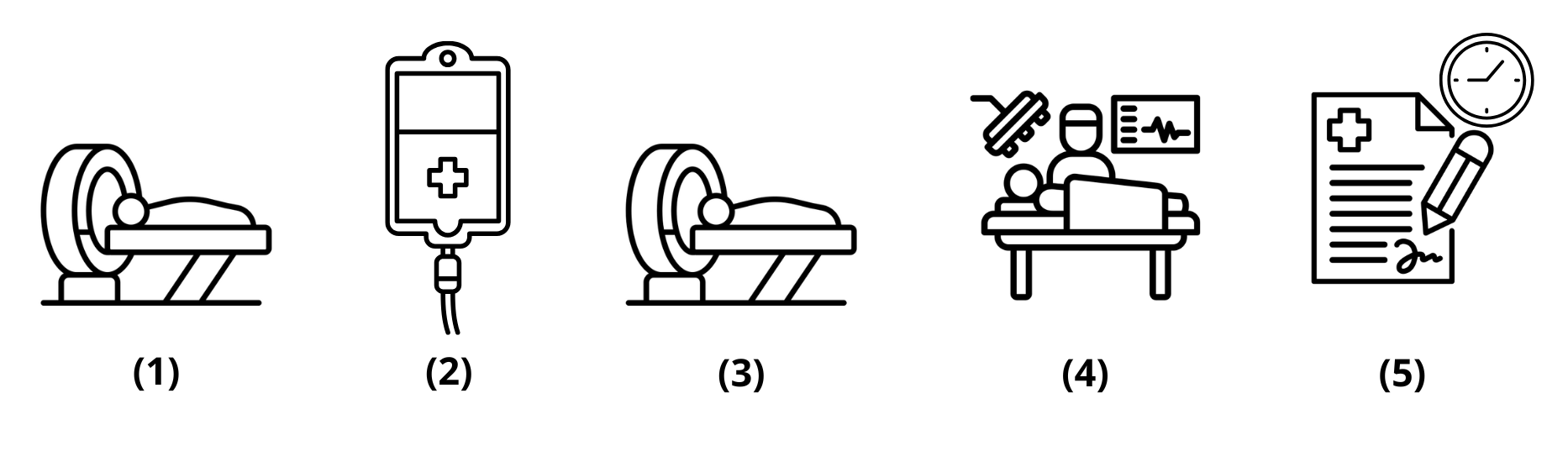}
    \caption{Schematic illustration of the clinical timeline for a representative patient in the cohort. (1) Baseline CT scan. (2) Chemotherapy. (3) Post-treatment CT scan. (4) Surgical resection. (5) Follow-up for relapse or death.}
    \label{fig:clinical_pathway}
\end{figure}


\subsection{Methodological pipeline}

This section outlines the methodological framework applied in our analysis, which comprises three main stages: (1) dataset preparation, including feature selection and transformation, necessary to construct a robust and interpretable input for downstream modelling; (2) model fitting; and (3) results interpretation.

\subsubsection{Dataset Preparation}

All baseline clinical variables were retained in the analysis. These include age, sex, number of metastases, presence of extrahepatic disease, and synchronous/metachronous metastasis. A small subset of patients (4 out of 102) had missing entries in one or more clinical variables; rather than apply imputation methods, which may introduce noise or bias given the low dimensionality and clinical semantics of the data, we excluded these patients. This resulted in a final cohort of 98 patients.

Radiomic feature selection was conducted to reduce redundancy and enhance interpretability. To ensure a longitudinal consistency, we performed the selection as follows. We separately selected $p$ variables in the baseline dataset \( \mathbf{X}^0 \in \mathbb{R}^{109} \) and $q$ variables in the post-treatment dataset \( \mathbf{X}^1 \in \mathbb{R}^{109} \), such that the two datasets resulted into \( \mathbf{X}^0 \in \mathbb{R}^p \) and \( \mathbf{X}^1 \in \mathbb{R}^q \) respectively. The selection was carried on based on pairwise Pearson correlation or Spearman correlation with a threshold of $0.9$.
Then, features belonging to the intersection \( p \cap q \), i.e., those with strong correlation in both imaging time points were discarded to avoid multi-collinearity and overfitting. This filtering step yielded 27 radiomic features for downstream analysis.

\begin{figure}[h!]
    \centering
    \includegraphics[width=0.85\textwidth]{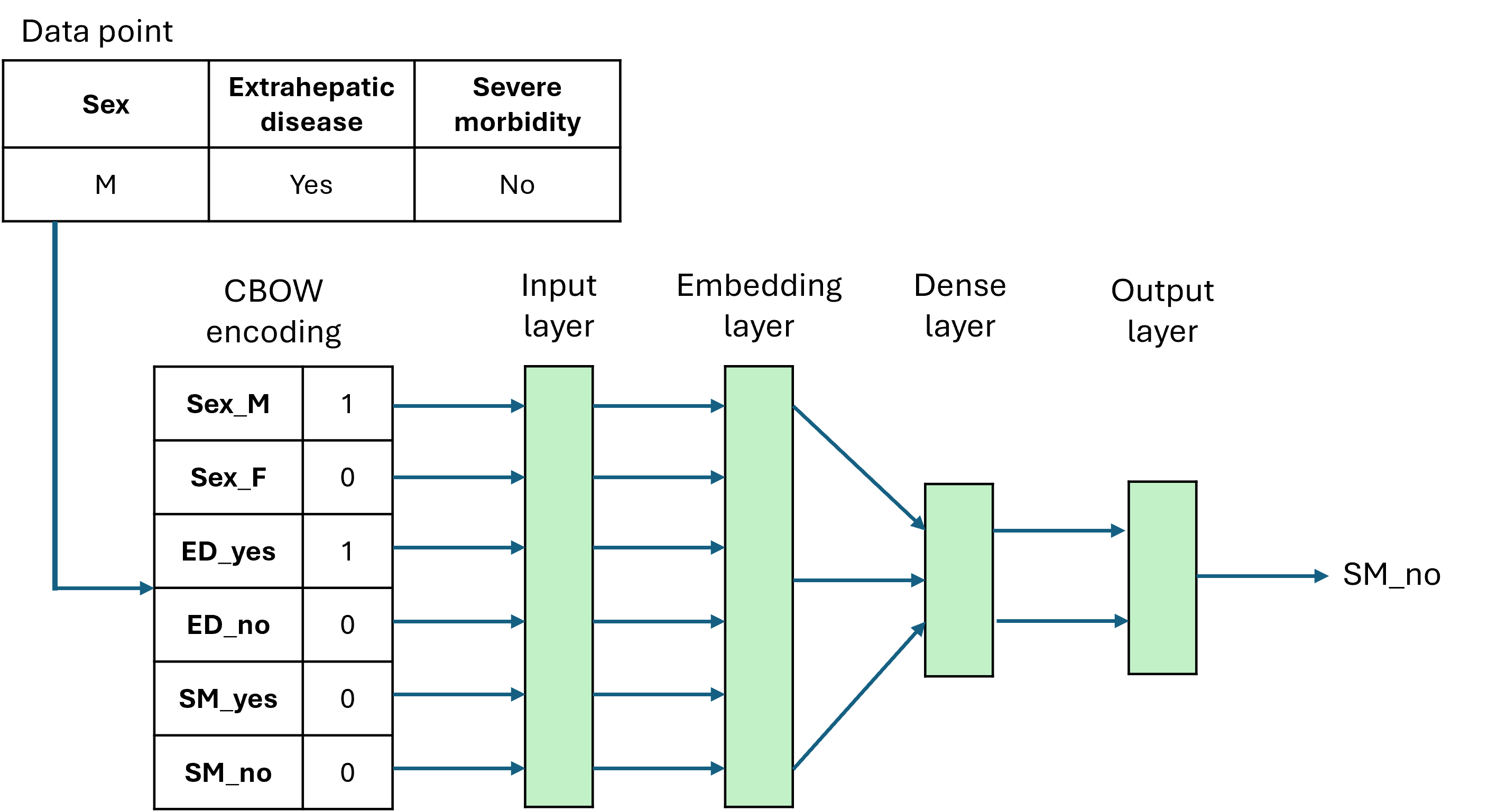}
    \caption{Schematic illustration of the embedding process for categorical variables using a Word2Vec-inspired Continuous Bag-of-Words model.}
    \label{fig:embedding}
\end{figure}

All numerical variables, including clinical and radiomic features, were standardized via z-score normalization while binary categorical variables (e.g., sex, metastasis timing, extrahepatic disease) were embedded in a 2-dimensional vector space to yield continuous, semantically informative representations. This embedding was learned using a variation of the \textit{Word2Vec} algorithm, adapted for categorical data \cite{waldemar2022word2vec}.
The core idea of this method parallels the \textit{Continuous Bag-of-Words} model in natural language processing. Each categorical variable value is treated analogously to a "word", and its "context" is defined by the co-occurrence with other categorical values across features for the same patient. A shallow neural network was trained to predict a target categorical value from its context, with an embedding layer mapping inputs to a dense vector space. The result was a set of learned 2D embeddings that capture latent similarities among categories beyond what is achievable via one-hot encoding.
The embedding framework is illustrated in Figure \ref{fig:embedding}. The final 2D embeddings are used as input features in place of their original categorical encodings.

\subsubsection{Model fitting and clustering}

\begin{figure}[h!]
    \centering
    \includegraphics[width=0.7\textwidth]{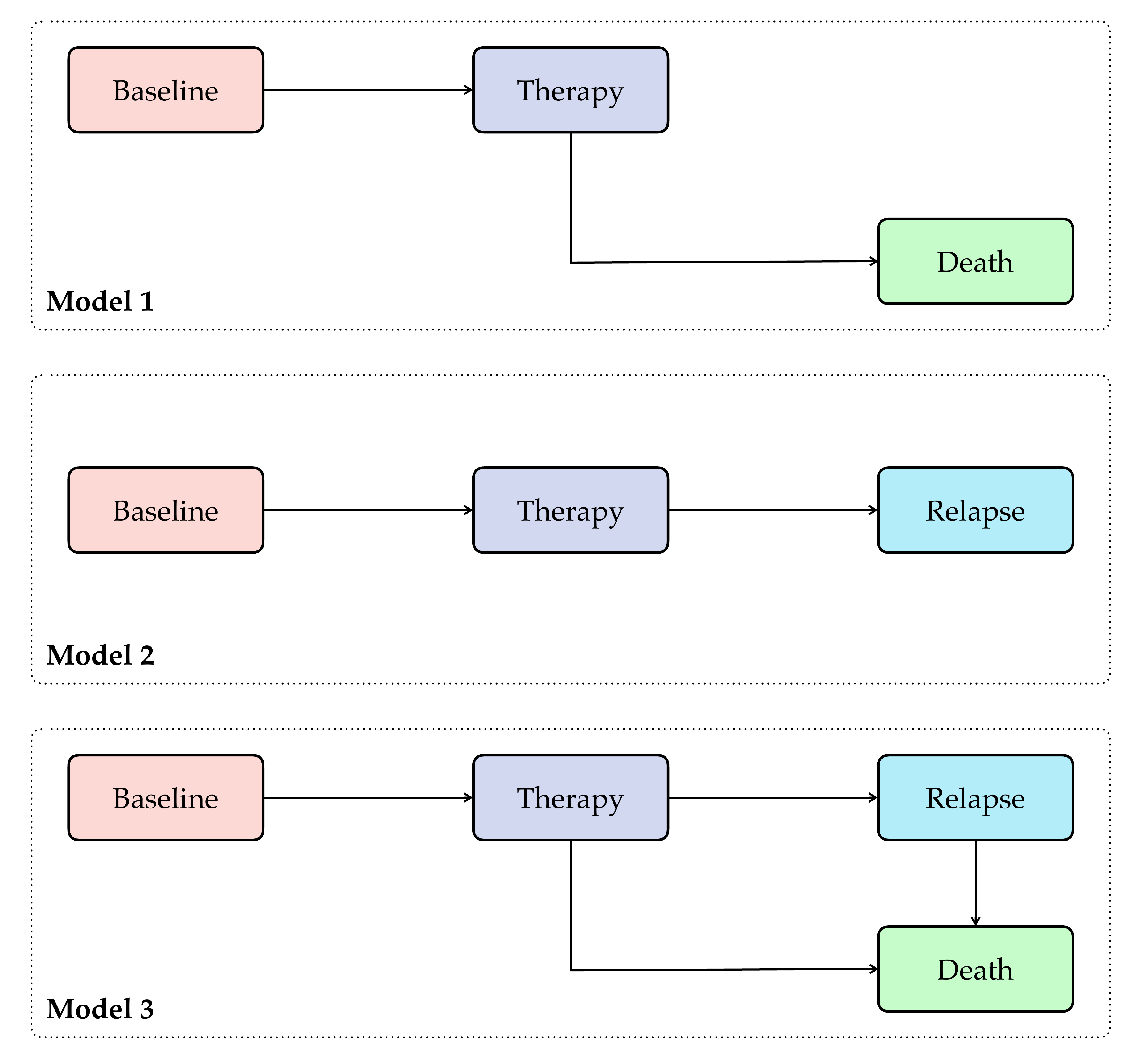}
    \caption{Schematic representation of the three clock-reset multi-state models (MSMs) used in this study. Each model encodes different assumptions about the timing and sequence of clinical transitions: Model 1 captures baseline to therapy to death; Model 2 includes relapse as an intermediate state before death; Model 3 allows for competing relapse vs death trajectories.}
    \label{fig:msm_models}
\end{figure}

To explore the latent structure of disease evolution and derive clinically meaningful subtypes, we defined three alternative clock-reset multi-state models (MSMs), each corresponding to a different assumption about transition timing and state dependency. These models, formally described and illustrated in Figure \ref{fig:msm_models}, specify distinct configurations of time-to-event relationships, transition mechanisms, and allowable state trajectories. All models are aligned with the clinical course of colorectal liver metastases, including systemic treatment, resection, relapse, and survival. 

Using each of these MSM configurations, we applied the trajectory-informed graph clustering approach detailed in the Section \ref{sec:model-formulation}. For a given model structure, the algorithm first estimates a patient similarity graph by jointly embedding clinical, radiomic, and time-to-event data within the corresponding MSM framework. Spectral clustering is then performed on the resulting graph Laplacian to identify patient subgroups that are, by design, similar in their longitudinal disease trajectories and baseline characteristics. This allowed us to derive coherent subgroups reflective of heterogeneous disease pathways.

Hyperparameter optimization was carried out via classical grid search. The tuning objective considered both the statistical relevance of the subtyping and its predictive quality. Specifically, we retained only those hyperparameter combinations that produced partitions of the cohort with significantly different survival distributions, as measured by a Log-Rank test with p-value $< 0.05$. Among these candidate configurations, we selected the optimal one based on the highest Harrell’s concordance index (C-index), which quantifies the discriminative power of the survival predictions.

\subsubsection{Evaluation of the goodness of fit}

To assess the quality and clinical relevance of the clustering results derived from each multi-state model (MSM), we employed an evaluation strategy involving both survival modelling and classification performance.

First, we evaluated the goodness-of-fit of the trajectory-informed model using Harrell’s concordance index (C-index), a standard metric for assessing the predictive accuracy of time-to-event models. The C-index was computed based on the estimated Cox proportional hazards weights \( \mathbf{\beta} \), quantifying the model’s ability to rank patients by risk across all possible pairwise comparisons.

Second, to test whether the derived subtypes are associated with differential disease trajectories, we fit a classical multi-state model using the \texttt{survival} and \texttt{mstate} packages in \texttt{R}. In this model, the only covariate was the cluster assignment label for each patient. A global Log-Rank test was applied to assess whether the survival curves across subgroups differed significantly. A p-value $< 0.05$ was considered evidence that the subtypes represent clinically distinct trajectories.

Third, for each discovered cluster \( C_k \), we trained a binary LASSO-regularized logistic regression model to classify whether an individual belongs to \( C_k \) or not and identify features associated with the labelling. The optimal hyperparameter \(\lambda\) was selected with a 5-fold cross validation monitoring the AUC metric. Classification performance was assessed using AUC in a 5-fold cross-validation scheme to ensure robustness. For those features that were significant at testing (p-value \(\leq\) 0.05), a violin plot was generated to highlight their difference between the two clusters.

Finally, to visualize the temporal dynamics of disease progression across subtypes, Kaplan-Meier (KM) curves were plotted for each clinically meaningful transition (e.g., from surgery to relapse or death), stratified by cluster label. These plots provide an interpretable depiction of subtype-specific risk profiles over time and highlight differences in progression patterns among the identified groups. Additionally, we used heatmaps to visualize transition probability - as \(1 - survival\ probability\) - within the multi-state models.


\subsection{Results}
We evaluated the trajectory-informed clustering method under three different clock-reset multi-state model (MSM) configurations. Each model captures a distinct structure of clinical state transitions, and clustering was performed using the patient similarity graphs estimated under each configuration. The evaluation includes predictive metrics including C-index, cluster discriminability via AUROC, and subgroup survival separation through Log-Rank tests and Kaplan-Meier (KM) curves. The aggregated summary of the results are available in Table \ref{tab:model_comparison} and, in the following, we detailed them.

\begin{table}[h!]
\centering
\begin{tabularx}{\textwidth}{l X c c c c}
\toprule
\textbf{Model} & \textbf{State Transitions} & \textbf{C-index} & \textbf{\# Clusters} & \textbf{Log-Rank p-value} & \textbf{AUC} \\
\midrule
Model 1 & Baseline $\rightarrow$ Therapy $\rightarrow$ Death & 0.67 & 2 & 0.04 & 0.74 \\
Model 2 & Baseline $\rightarrow$ Therapy $\rightarrow$ Relapse & 0.60 & 2 & 0.05 & 0.94 \\
Model 3 & Baseline $\rightarrow$ Therapy $\rightarrow$ \{Relapse, Death\} & 0.64 & 2 & 0.003 & 0.87 \\
\bottomrule
\end{tabularx}
\caption{Comparison of the three multi-state models in terms of predictive performance, survival stratification, and clinical relevance.}
\vspace{1em}

\begin{tabularx}{\textwidth}{l X}
\toprule
\textbf{Model} & \textbf{Kaplan-Meier Insights and Clinical Utility} \\
\midrule
Model 1 & Strong survival curve differentiation; red cluster shows survival predominantly within 5 years, while blue cluster has a 50\% probability of surviving up to 10 years. Separation suggests some prognostic relevance and clinical utility for precise risk stratification.\\
Model 2 & Minimal relapse-based separation; survival curves are almost overlapped. High AUC may indicate feature-driven similarity rather than survival-based stratification. Limited clinical utility.\\
Model 3 & Meaningful separation in relapse risk; differentiation in transition from therapy to death, though limited by low event frequency. No meaningful stratification detected in relapse-to-death transition. Clinically useful for for guiding treatment intensity and monitoring strategies based on relapse risk, but provides limited insight into post-relapse survival outcomes. \\
\bottomrule
\end{tabularx}
\label{tab:model_comparison}
\end{table}

\paragraph{Model 1: Progression to Death}

Model 1 represents a progressive three-state pathway capturing the natural sequence of events from diagnosis to therapy and, ultimately, to death. Patients begin in a baseline state, transition to a post-chemotherapy state, and may subsequently reach an absorbing death state following surgery.

The model was configured with parameter settings: \(\eta = 0.0001\), \(\gamma = 10000\), \(\mu = 0.00005\), \(k = 7\). At the last iteration of the optimization procedure, the estimated transition weights based on the inverse of the parameters variance are 2.04 for the transition \textit{Baseline} $\rightarrow$ \textit{Therapy} and 2.32 for the transition \textit{Therapy} $\rightarrow$ \textit{Death}. Training dynamics for the model, including the evolution of the loss function across epochs, are provided in the Supplementary Materials.

Upon convergence, the model achieved a C-index of 0.67, indicating moderate accuracy in survival prediction. Spectral clustering applied to the similarity matrix resulted in 2 patient subgroups. Stratification by cluster in a survival MSM yielded a Log-Rank test p-value of 0.04, suggesting statistically significant—but modest—differences in overall survival. The classifier distinguishing cluster membership based on clinical and radiomic features achieved a AUC of 0.74.

\begin{figure}[h!]
    \centering
    \includegraphics[width=0.4\textwidth]{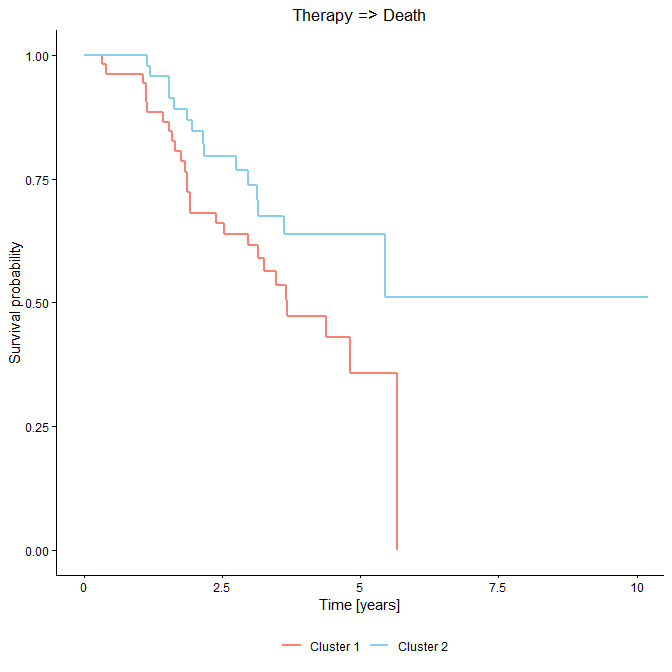}
    \caption{Kaplan-Meier curve for Model 1 (therapy to death). The two clusters show significant separation, especially from year 5 onwards.}
    \label{fig:km_model1}
\end{figure}

The Kaplan-Meier curves (Figure \ref{fig:km_model1}) demonstrate a clear and substantial separation between the two clusters, indicating meaningful differences in survival outcomes. Patients in the blue cluster exhibit significantly better survival probability compared to those in the red cluster, with the curves diverging notably from approximately 1 year onwards and the gap widening progressively over time. While the red cluster shows a relatively steep decline in survival, with the majority of patients experiencing the event of death by around 6 years, the blue cluster presents a much more favorable prognosis. This group maintains a stable survival probability of approximately 0.50 after 5 years, with this proportion of patients remaining alive and event-free extending to the end of the 10-year observation period. This strong and sustained separation throughout the follow-up period strongly suggests that the underlying clustering effectively captures significant and clinically relevant heterogeneity in patient trajectories with respect to survival outcomes, providing valuable insights for differentiated risk assessment.  The distinct long-term survival patterns imply that the two identified clusters represent genuinely different risk groups within the patient population, underscoring the potential for tailored therapeutic approaches. This clear separation highlights a clinically actionable distinction: the model effectively isolates a high-risk group that may benefit from intensified monitoring or adjuvant therapy, as well as a lower-risk group that could potentially be spared unnecessary interventions.

\begin{figure}[h!]
    \centering
    \includegraphics[width=1\textwidth]{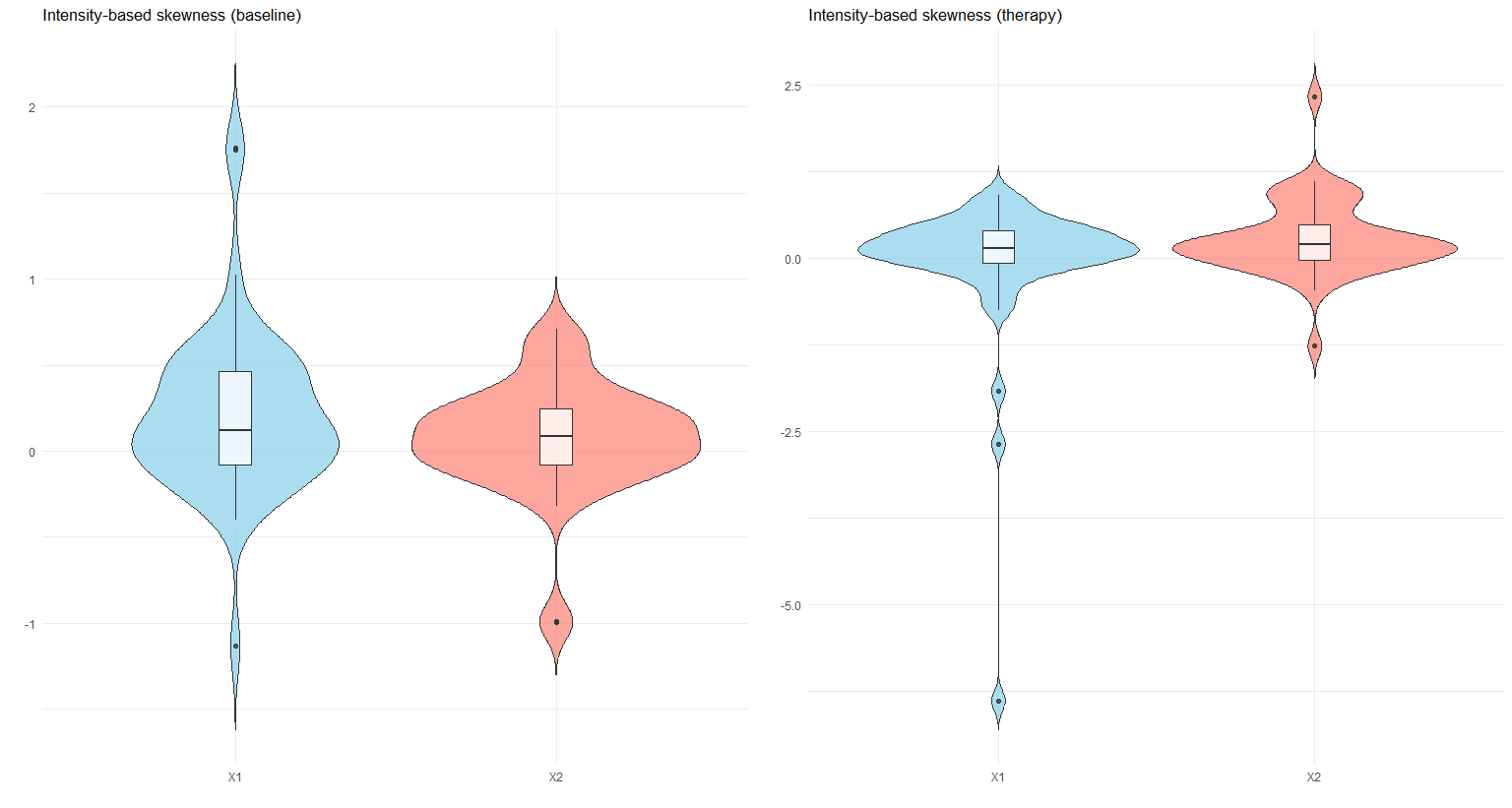}
    \caption{Violin plots for significant features in Model 1: Intensity-based skewness at baseline and after chemotherapy.}
    \label{fig:bp_model1}
\end{figure}

Intensity-based skewness, both before and after chemotherapy, stands out as the only radiomic feature that achieves statistical significance (p-value \(<\) 0.05) in predicting the cluster a patient belongs to, when survival data are not included in the classification task (Figure \ref{fig:bp_model1}).

\begin{figure}[h!]
    \centering
    \includegraphics[width=1\textwidth]{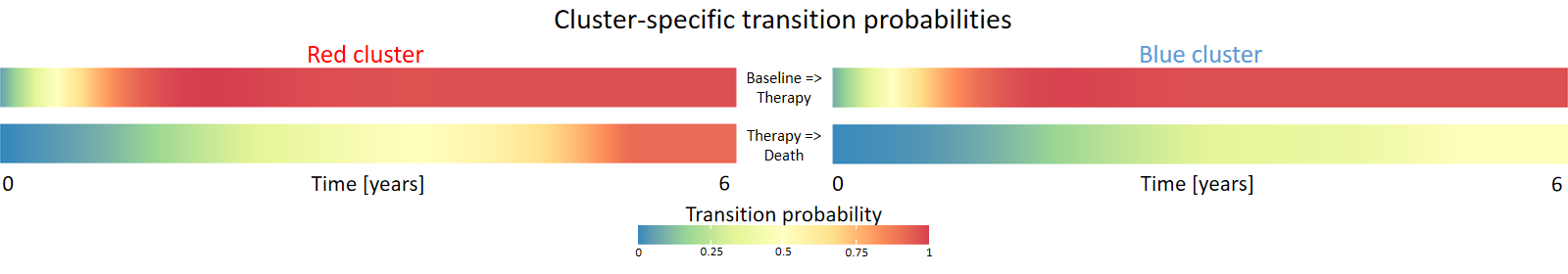}
    \caption{Graphical representation of transition probabilities for all transitions over time in the multi-state model 1. The red cluster shows a higher transition probability toward death than the blue cluster around four years after therapy.}
    \label{fig:pathway_model1}
\end{figure}

Figure \ref{fig:pathway_model1} displays a graphical comparison between the transition probabilities of an individual based on the cluster to which a patient has been assigned to. It is possible to notice a clear difference between the two clusters in the transition to the death state.

\paragraph{Model 2: Progression to Relapse}

Model 2 adopts a similar structure but redefines the final absorbing state as disease relapse rather than death. This change emphasizes early disease recurrence as a clinically meaningful event distinct from mortality, and may better reflect the course of metastatic colorectal cancer, where relapse is often a pivotal moment in disease management.

Parameters for this model were set as: \(\eta = 0.001\), \(\gamma = 1000\), \(\mu = 0.001\), \(k = 6\). Transition weights after the last iteration of the optimization algorithm are 2.35 for the transition \textit{Baseline} $\rightarrow$ \textit{Therapy} and 2.62 for the transition \textit{Therapy} $\rightarrow$ \textit{Relapse}. The final C-index was 0.60, and the clustering again yielded two patient subgroups. The clusters showed moderate survival differentiation, with a Log-Rank p-value of 0.05 and an AUC of 0.94 for the classifier.

\begin{figure}[h!]
    \centering
    \includegraphics[width=0.4\textwidth]{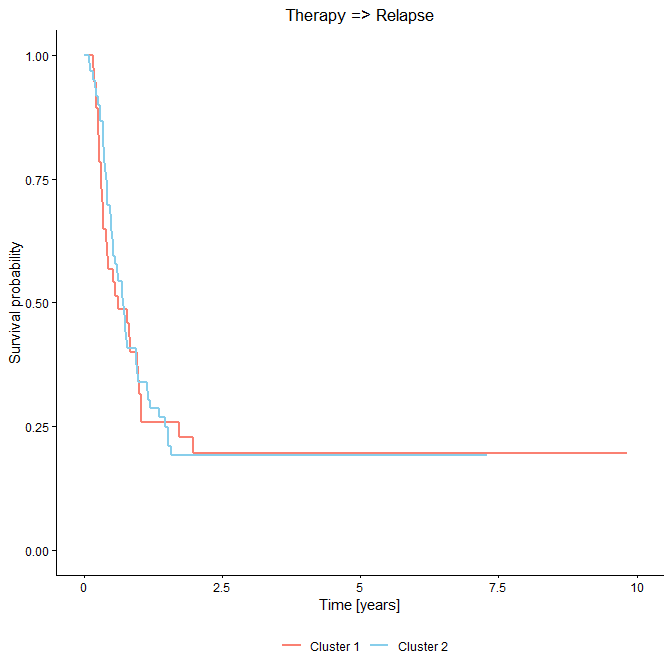}
    \caption{Kaplan-Meier curve for Model 2 (therapy to relapse). The two curves show a similar progression over time}
    \label{fig:km_model2}
\end{figure}

Despite the moderate p-value obtained from the Log-Rank test p-value and the high accuracy achieved by the classifier in predicting patient clusters, the inspection of the Kaplan-Meier curves (Figure \ref{fig:km_model2}) do not reveal a clear survival stratification between the groups with respect to the event of disease relapse. The survival curves show overlap and only a slight separation, indicating that the clustering does not strongly differentiate patient outcomes based on survival. This discrepancy suggests that the classifier’s strong performance may be driven primarily by the radiomic and clinical features used for clustering, rather than by meaningfully integrting survival information: the clustering task appears to rely heavily on feature similarity without adequately capturing the heterogeneity in survival trajectories, which limits its ability to produce clinically relevant survival-based stratification.

\begin{figure}[h!]
    \centering
    \includegraphics[width=1\textwidth]{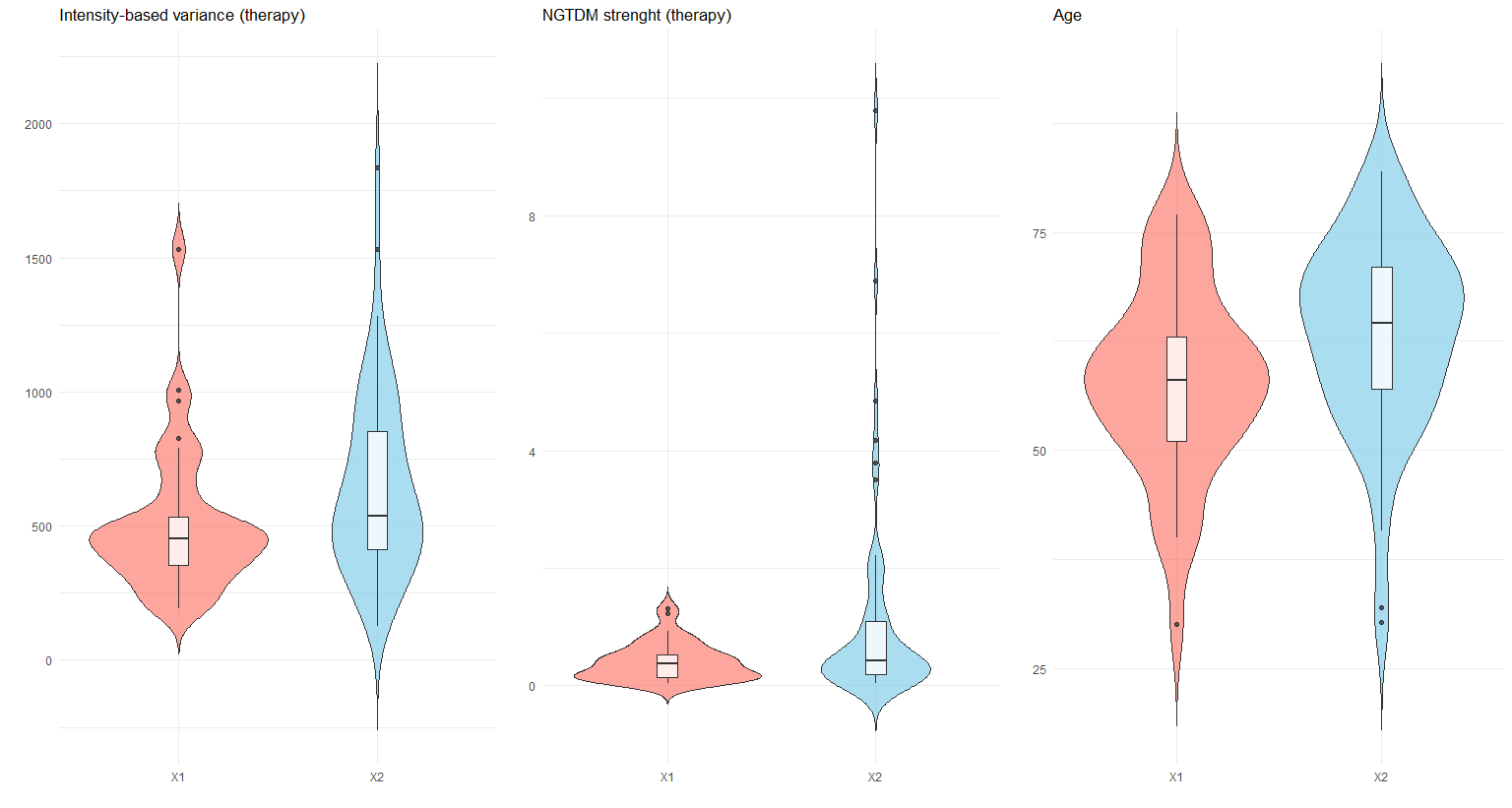}
    \caption{Violin plots for significant features in Model 2: Intensity-based variance after chemotherapy, Neighbouring Gray Tone Difference Matrix strength after chemotherapy and age.}
    \label{fig:bp_model2}
\end{figure}

The significant features identified for predicting an individual’s cluster based on radiomic and clinical data are Intensity-based variance after chemotherapy, Neighbouring Gray Tone Difference Matrix (NGTDM) strength after chemotherapy, and age (Figure \ref{fig:bp_model2}).

\begin{figure}[h!]
    \centering
    \includegraphics[width=1\textwidth]{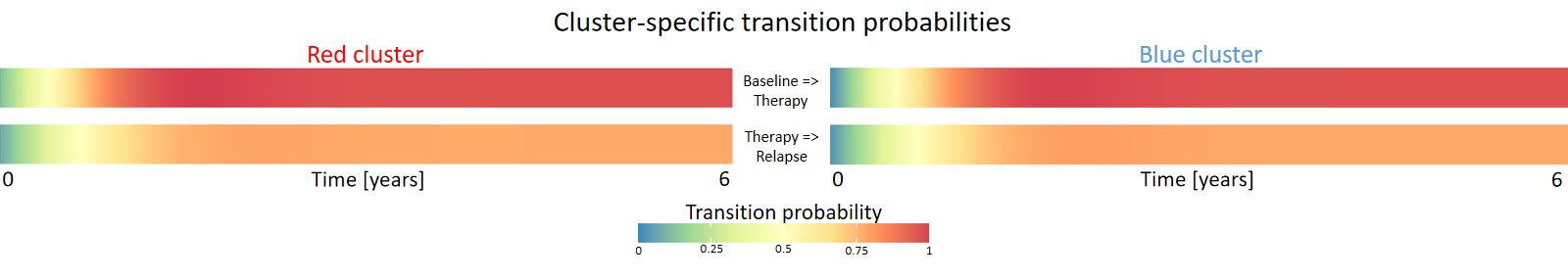}
    \caption{Graphical representation of transition probabilities for all transitions over time in the multi-state model 2. The two clusters show very similar transition probabilities over the timeline}
    \label{fig:pathway_model2}
\end{figure}

Figure \ref{fig:pathway_model2} displays a graphical comparison between the transition probabilities of an individual based on the cluster to which a patient has been assigned to. As in the Kaplan-Meier plots, a clear differentiation between the two clusters is not achieved.

\paragraph{Model 3: Competing Risks of Relapse and Death}

Model 3 introduces a more nuanced clinical picture by simultaneously modelling both relapse and death as competing absorbing states. This accounts for the non-trivial portion of patients who die without experiencing relapse, offering a more complete view of patient trajectories.

With parameters set to \(\eta = 0.001\), \(\gamma = 1000\), \(\mu = 0.0005\), \(k = 6\) and weights at the end of the optimization process set to 1.82 for the transition \textit{Baseline} $\rightarrow$ \textit{Therapy}, 1.54 for the transition \textit{Therapy} $\rightarrow$ \textit{Death}, 2.54 for the transition \textit{Therapy} $\rightarrow$ \textit{Relapse} and 2.08 for the transition \textit{Relapse} $\rightarrow$ \textit{Death}, this model achieved the C-index of 0.64. Spectral clustering produced two distinct subgroups, with the best Log-Rank test p-value of 0.003 and an AUC of 0.87- lower than Models 2, but reflecting the added complexity of modelling multiple outcomes.

The fact that transition \textit{Therapy} $\rightarrow$ \textit{Death} has the lowest weight, and consequently the highest variance in the parameters associated with that transition, is expected given that only six patients in the dataset experience this transition. This limited sample size results in less information to reliably estimate the parameters, leading to greater uncertainty and variability. 

\begin{figure}[h!]
    \centering
    \includegraphics[width=\textwidth]{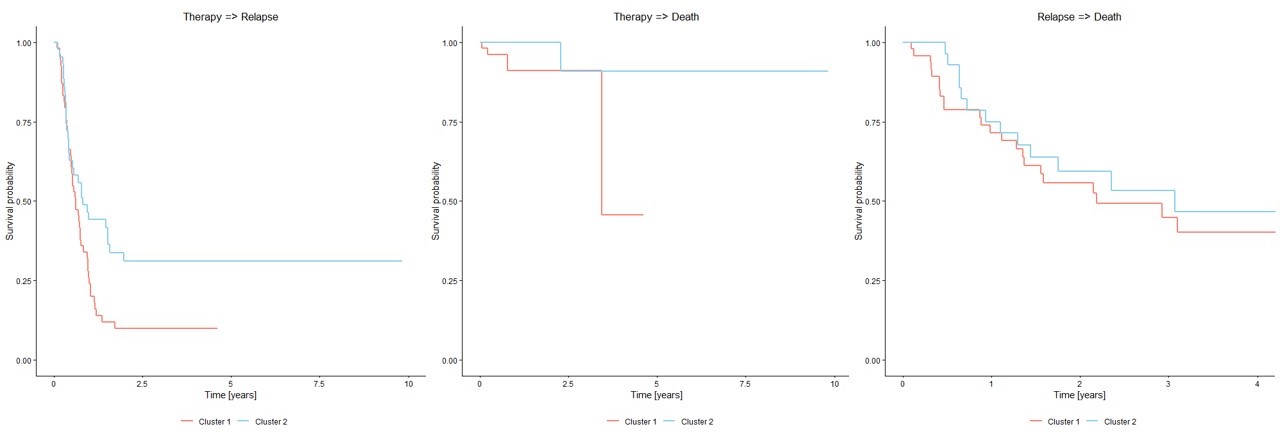}
    \caption{Kaplan-Meier curve for Model 3: a) therapy to relapse; b) therapy to death; c) relapse to death.}
    \label{fig:km_model3}
\end{figure}

The KM curve for the transition from therapy to relapse (Figure The Kaplan-Meier curve for the transition from therapy to relapse (Figure \ref{fig:km_model3}\textit{a}) shows meaningful subgroup differentiation. The red cluster exhibits the worst prognosis, with rapid and early relapse. The blue cluster initially demonstrates a survival trend similar to the red cluster, though it diverges significantly after approximately 0.5 to 1 year, maintaining a higher relapse-free survival probability and indicating that some individuals reach a longer disease-free period.

The transition from therapy to death (Figure \ref{fig:km_model3}\textit{b}) suffers the same problem as described before about the lack of a sufficient amount of data.

The transition from relapse to death (Figure \ref{fig:km_model3}\textit{c}) produced curves that, while showing some periods of overlap, also demonstrate a consistent trend of differentiation. The blue cluster generally maintains a better survival probability than the red cluster, with the curves beginning to separate more clearly after approximately 1.5 years. However, a strong differentiation is not noticeable, differently from the Baseline -> Relapse transition. This may suggest that once relapse occurs, survival is influenced by factors that are partially captured by the current features or model structure, with the blue cluster still exhibiting a survival advantage.

The similarity of the KM curves for the relapse-to-death transition across clusters may suggest that once a patient enters the relapse stage, there is minimal differentiation in the probability of death. This indicates that the current model and features may not capture factors influencing survival post-relapse, highlighting a relative homogeneity in outcomes after relapse occurs. This is clinically plausible, as treatment guidelines often consider relapse to represent essentially a new disease state, prompting an entirely different therapeutic pathway that overrides prior risk distinctions. Based on the observed similarity in survival curves after relapse, it is possible to hypothesize that interventions aimed at preventing relapse might have a greater impact on patient survival than efforts focused on stratifying risk post-relapse. While this interpretation is supported by the current data, further clinical investigation is needed to confirm whether prioritizing relapse prevention indeed leads to improved outcomes.

\begin{figure}[h!]
    \centering
    \includegraphics[width=1\textwidth]{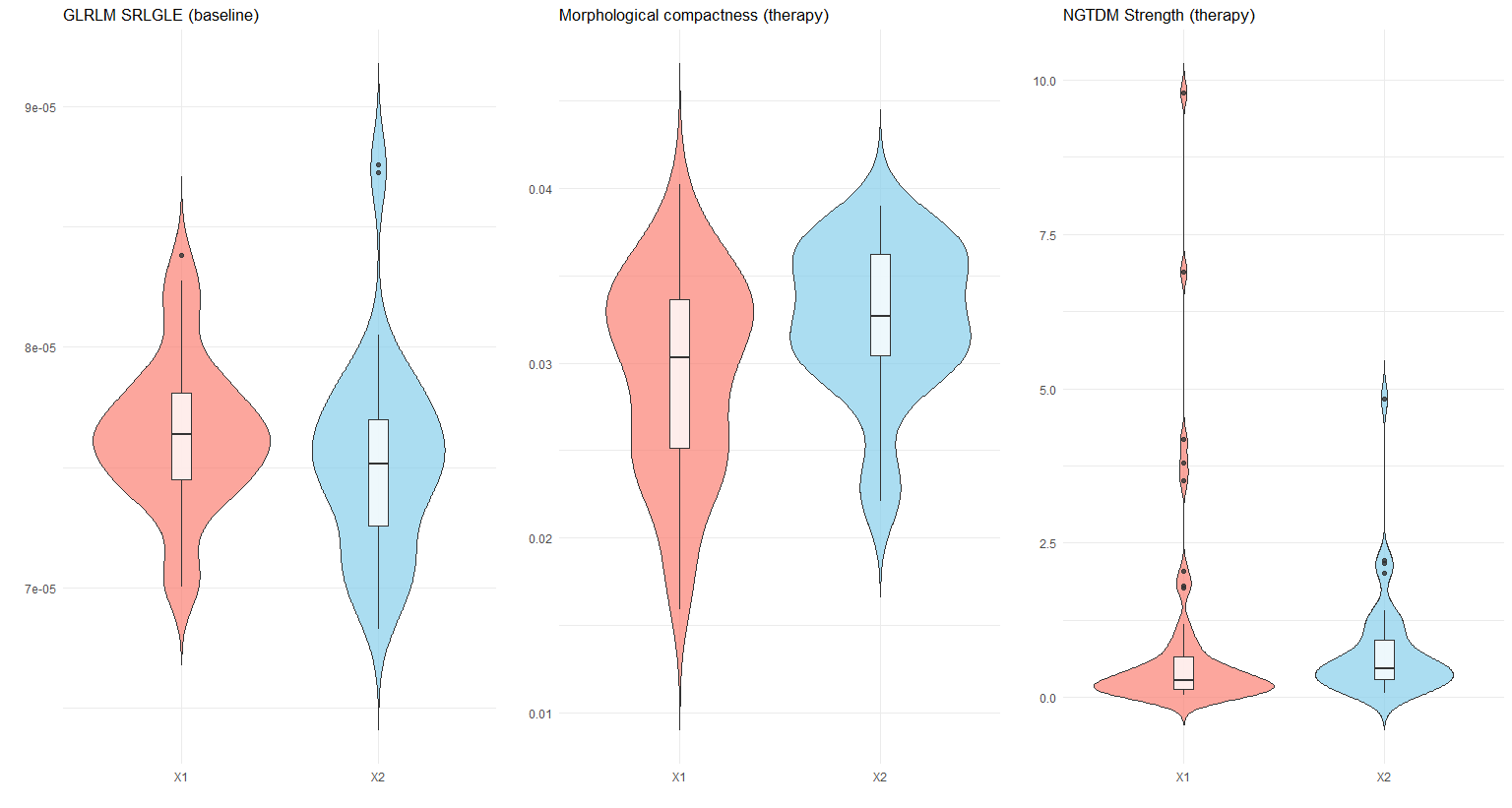}
    \caption{Violin plots for significant features in Model 3: Grey-Level Run Length Matrix Short Run Low Grey Level Emphasis before chemotherapy, Morphological Compactness after chemotherapy and Neighbouring Gray Tone Difference Matrix Strength after chemotherapy.}
    \label{fig:bp_model3}
\end{figure}

The features Grey-Level Run Length Matrix Short Run Low Grey Level Emphasis before chemotherapy, Morphological Compactness after chemotherapy, and Neighbouring Gray Tone Difference Matrix Strength after chemotherapy were found to be significant in the fitted classifier for predicting the cluster to which a patient belongs (Figure \ref{fig:bp_model3}).

\begin{figure}[h!]
    \centering
    \includegraphics[width=1\textwidth]{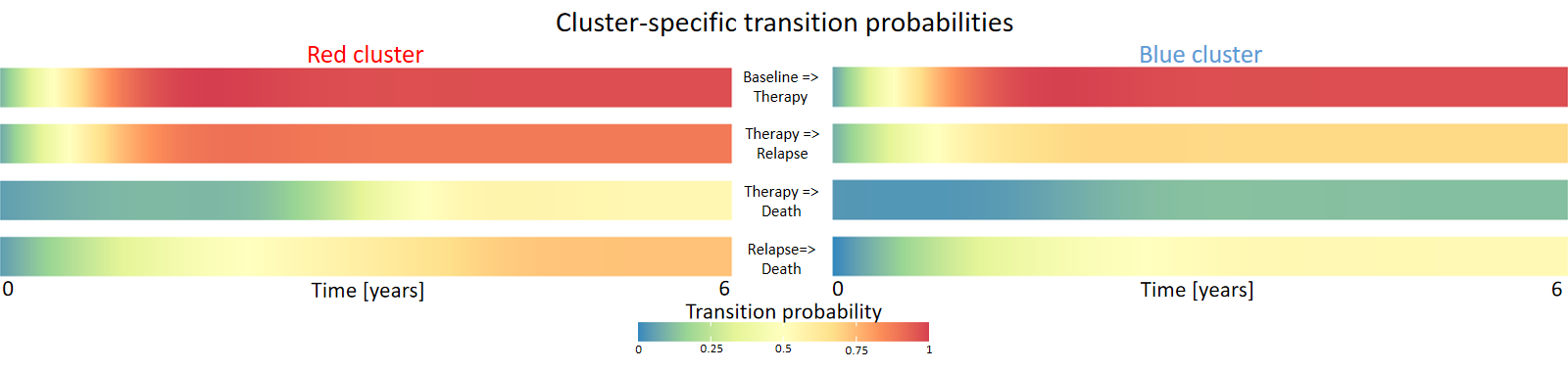}
    \caption{Graphical representation of transition probabilities for all transitions over time in the multi-state model 3. The two clusters show very significant differentiation in all transition probabilities over the timeline}
    \label{fig:pathway_model3}
\end{figure}

Figure \ref{fig:pathway_model3} displays a graphical comparison between the transition probabilities of an individual based on the cluster to which a patient has been assigned to. It is possible to see that all the transition to events of interest present a slight-to-moderate, but yet noticeable, differentiation in the two clusters.

\subsection{Discussion}

Among the three models evaluated, each demonstrates distinct strengths and limitations in terms of survival prediction performance, possibility of determine the cluster to which a patient belongs to a-priori and clinical relevance. 

Model 1 achieves the highest concordance index (0.67), indicating it is the most effective at predicting patients' survival. The Kaplan-Meier curves for this model show a clear ability to stratify individuals by death risk, suggesting clinical utility in identifying high-risk patients. 

Model 2 is the one with the highest area under the curve (AUC), reflecting strong accuracy in predicting cluster membership based on radiomic and clinical features. Despite this, the analysis of the KM plot reveals that Model 2 has very low discrimination capability with respect to the event of disease relapse, implying that its high AUC may be driven more by the similarity of input features rather than by meaningful differences in patient outcomes. This raises concerns that the model may overfit to feature patterns without effectively capturing clinically relevant survival heterogeneity. 

Model 3, despite its increased complexity, presents a compelling trade-off between predictive power and the ability to differentiate survival outcomes with respect to both relapse and death events. While achieving intermediate C-index and AUC values, its most notable strength lies in generating the most significant Log-Rank test p-value, which translates to a more pronounced separation of survival curves among all possible transitions. Model 3 demonstrates an improved capacity to stratify patients by relapse risk compared to model 2, providing a more meaningful distinction between subgroups for this outcome. However, this comes at the cost of a less precise stratification for death risk. This suggests that while incorporating competing risks and additional clinical information may offer a better overall view of patient trajectories by accounting for multiple possible events, it can lead to a loss of precision when specifically stratifying the risk for individual outcomes like death.
Overall, Model 3’s increased complexity allows it to capture aspects of patient heterogeneity that the other models may miss, making it a reasonable compromise between predictive accuracy and clinical interpretability. 
Models 1 and 2 demonstrate strengths in survival ranking and feature-based classification, respectively, but their limitations highlight the challenges of achieving both high predictive performance and meaningful clinical stratification in heterogeneous patient populations.

Across all three models, our trajectory-informed clustering approach demonstrated the ability to identify subgroups with differing survival patterns. While quantitative metrics alone were comparable, inspection of Kaplan-Meier curves was essential to uncover meaningful stratification. Notably, Model 3 achieved the best balance between interpretability, predictive accuracy and clinical utility when focusing on the death event. This reinforces the promise of integrating radiomic features, clinical data, and temporal modelling to uncover subtypes that are not only statistically distinct, but also actionable in guiding therapy, patient stratification, and future clinical trial design.

\section{Conclusions}
\label{sec:conclusions} 
In this work, we proposed a novel approach for cancer subtyping based on longitudinal data. We first evaluated the model on a set of randomly generated datasets comparing the results with other methods and then performed an ablation study to assure that every component of the optimization problem has a positive impact on the final result. Finally, we applied the method to real-world dataset of patients affected by colorectal liver metastases employing three different multi-state models. Two out of three models demonstrated promising results in identifying two distinct subgroups of patients, each with different disease-free and survival time expectations. Notably, the algorithm could be leveraged to enhance clinical decision-making and ultimately improve patient outcomes by tailoring treatment planning.\\
While this work presents a valuable method for cancer subtyping, there are two key limitations that warrant attention in future research. First, the approach assumes Markovian transitions in the multi-state model, which restricts its ability to incorporate variations in covariates between states. Second, the method’s representation of competing risks through a linear combination, though flexible, may
oversimplify the complexities of patient heterogeneity in the presence of competing risks. Addressing these limitations could enhance the model’s robustness and its applicability in clinical settings.

\newpage

\section*{Supplementary Materials}
\appendix

\subsection*{Patients' clinical description}

\begin{table}[ht]
    \centering
    \caption*{\textbf{A. Patients' Outcomes}}
    \begin{tabularx}{\linewidth}{Xcc}
        \toprule
        \textbf{Feature} & \textbf{Frequency} & \textbf{Percentage} \\
        \midrule
        Disease free & 17 & 17.3\% \\
        Alive with disease relapse & 39 & 39.8\% \\
        Dead after disease relapse & 37 & 37.8\% \\
        Dead without disease relapse & 5 & 5.1\% \\
        \bottomrule
    \end{tabularx}

    \vspace{1em}

    \caption*{\textbf{B. Patients' Clinical Categorical Features}}
    \begin{tabularx}{\linewidth}{Xcc}
        \toprule
        \textbf{Feature} & \textbf{Values} & \textbf{Percentage} \\
        \midrule
        \multirow{2}{=}{Sex} 
        & M: 59 & M: 60.3\% \\
        & F: 39 & F: 39.7\% \\
        \midrule
        \multirow{2}{=}{Extrahepatic disease} 
        & Yes: 22 & Yes: 22.5\% \\
        & No: 76 & No: 77.5\% \\
        \midrule
        \multirow{2}{=}{Synchronous/Metachronous} 
        & S: 72 & S: 73.4\% \\
        & M: 26 & M: 26.6\% \\
        \bottomrule
    \end{tabularx}

    \vspace{1em}

    \caption*{\textbf{C. Patients' Clinical Numerical Features}}
    \begin{tabularx}{\linewidth}{Xccc}
        \toprule
        \textbf{Feature} & \textbf{Mean} & \textbf{SD} & \textbf{Median} \\
        \midrule
        Age & 61.3 & 11.1 & 61.5 \\
        N metastases & 6.1 & 4.4 & 5.0 \\
        \bottomrule
    \end{tabularx}
    \caption{Patient Clinical Data Summary}

\end{table}

\newpage

\subsection*{Patients' radiomics description}

\begin{table}[htbp]
\centering
\small
\begin{tabularx}{\linewidth}{l *{6}{>{\raggedleft\arraybackslash}X}}
\toprule
\textbf{Feature} & 
\multicolumn{3}{c}{\textbf{Radiomics baseline}} & 
\multicolumn{3}{c}{\textbf{Radiomics post-treatment}} \\
\cmidrule(lr){2-4} \cmidrule(lr){5-7}
& Mean & SD & Median & Mean & SD & Median \\
\midrule
MORPHOLOGICAL\_Volume & 69428 & 142938 & 16918 & 18373 & 50899 & 4437 \\
MORPHOLOGICAL\_ApproximateVolume & 69537 & 143016 & 17031 & 18445 & 50963 & 4489 \\
MORPHOLOGICAL\_Compactness & 0.02897 & 0.00568 & 0.02963 & 0.03075 & 0.005366 & 0.031875 \\
MORPHOLOGICAL\_CentreOfMassShift & 0.14790 & 0.13184 & 0.10568 & 0.47111 & 0.39687 & 0.36119 \\
INTENSITY-BASED\_Mean & 69.5570 & 18.3011 & 69.1854 & 73.710 & 20.975 & 70.527 \\
INTENSITY-BASED\_Variance & 646.842 & 980.291 & 507.205 & 593.13 & 319.05 & 503.22 \\
INTENSITY-BASED\_Skewness & 0.14812 & 0.41450 & 0.11977 & 0.10160 & 0.87914 & 0.15893 \\
INTENSITY-BASED\_Kurtosis & 0.90497 & 4.19676 & 0.08374 & 1.58793 & 10.241 & 0.04260 \\
INTENSITY-BASED\_MinimumGreyLevel & -34.557 & 125.296 & -10.5 & -8.7244 & 74.4167 & 1.0 \\
INTENSITY-BASED\_90thPercentile & 100.076 & 21.3577 & 98.5 & 104.13 & 23.340 & 101.5 \\
INTENSITY-BASED\_MaximumGreyLevel & 182.773 & 102.811 & 161 & 165.417 & 55.4146 & 148.5 \\
INTENSITY-BASED\_InterquartileRange & 31.3434 & 11.5948 & 30.5 & 31.6734 & 9.37178 & 30.0 \\
INTENSITY-BASED\_Range & 217.331 & 200.652 & 167.5 & 174.142 & 99.0992 & 151.5 \\
INTENSITY-BASED\_CoefficientOfVariation & 0.37166 & 0.20153 & 0.33595 & 0.344205 & 0.131670 & 0.315178 \\
GLCM\_JointMaximum & 0.04824 & 0.02229 & 0.04264 & 0.050196 & 0.02248 & 0.045354 \\
GLCM\_DifferenceAverage & 1.84829 & 0.94464 & 1.72193 & 1.85608 & 0.45142 & 1.77284 \\
GLCM\_DifferenceEntropy & 4.50187 & 2.72321 & 5.80116 & 6.08266 & 0.61347 & 6.12462 \\
GLCM\_NormalisedInverseDifference & 0.91651 & 0.02641 & 0.91962 & 0.90518 & 0.026365 & 0.90658 \\
GLCM\_Correlation & 0.43917 & 0.17071 & 0.44016 & 0.40512 & 0.153657 & 0.425651 \\
GLCM\_ClusterShade & -19.497 & 318.784 & 4.44652 & 18.5405 & 67.2203 & 10.0526 \\
GLRLM\_ShortRunsEmphasis & 0.87600 & 0.03344 & 0.87920 & 0.89319 & 0.02712 & 0.89417 \\
GLRLM\_ShortRunLowGreyLevelEmphasis & 7.58e-5 & 3.58e-6 & 7.59e-5 & 7.67e-5 & 3.00e-6 & 7.68e-5 \\
GLRLM\_ShortRunHighGreyLevelEmphasis & 10155 & 577.530 & 10145 & 10435.9 & 603.289 & 10347.3 \\
NGTDM\_Contrast & 0.04572 & 0.02652 & 0.04194 & 0.06325 & 0.03761 & 0.05396 \\
NGTDM\_Strength & 0.22044 & 0.36028 & 0.12711 & 0.82447 & 1.14152 & 0.40177 \\
GLSZM\_SmallZoneEmphasis & 0.59157 & 0.06448 & 0.59989 & 0.57793 & 0.05305 & 0.5909 \\
GLSZM\_ZoneSizeEntropy & 5.73740 & 0.53856 & 5.81217 & 5.48960 & 0.60256 & 5.56010 \\
\bottomrule
\end{tabularx}
\caption{Radiomics Features: Baseline vs Post-treatment}

\end{table}

\newpage

\subsection*{Training Loss Curves}
This section presents the loss curves of the proposed algorithm applied to the presented case study. At each iteration of the outermost loop of Algorithm \ref{algoritmo}, the total loss function was evaluated using the current estimates of \(S\) and \(\beta\). These loss values were recorded at each iteration to monitor the progression of the optimization process. Since the initialization values for both \(S\) and \(\beta\) are fixed and no stochastic elements were introduced, the entire procedure is deterministic, ensuring reproducibility of the results. After completing all iterations, a loss graph was generated by plotting the total loss against the iteration number.

\begin{figure}[ht]
    \centering
    \includegraphics[width=\linewidth]{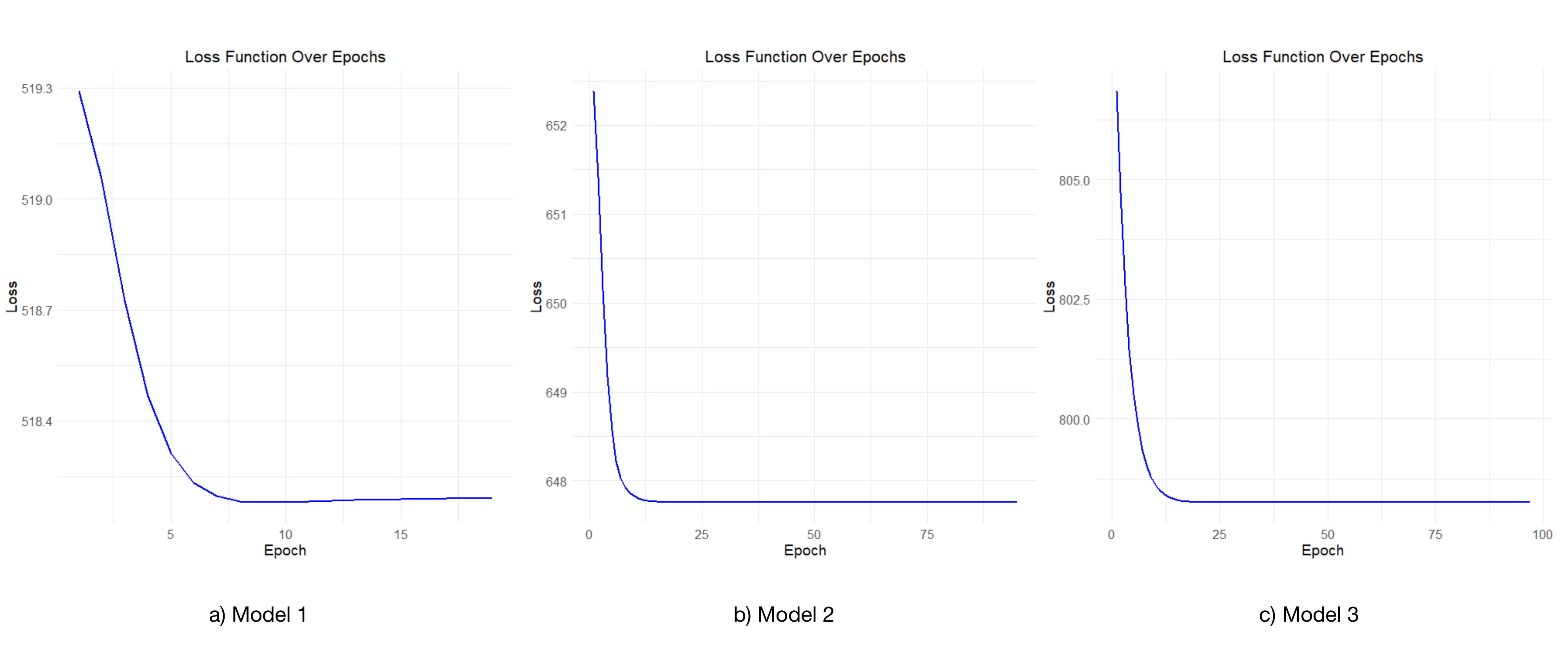}
    \caption{Training losses for the three fitted models, each reflecting different assumptions about clinical transition dynamics. Model 1 represents a linear progression from baseline to therapy and ultimately to death. Model 2 incorporates relapse as an intermediate state preceding death. Model 3 introduces competing trajectories, allowing patients to either relapse or die following therapy. Panels a), b), and c) correspond to Models 1, 2, and 3, respectively.}
    \label{fig:training}
\end{figure}

\newpage

\section*{Acknowledgements}
The authors wish to thank the Italian Association for Cancer Research (AIRC - grant \#2019-23822), the physicians who contributed to data collection and the patients. 
L. Cavinato is funded by the National Plan for NRRP Complementary Investments “Advanced Technologies for Human-centred Medicine” (PNC0000003). The present research is part of the activities of “Dipartimento di Eccellenza 2023-2027".

\bibliographystyle{unsrt}
\bibliography{sample}

@online{cancer_who,
  author={{World Health Organization}},
  title={Cancer},
  year=2024,
  url={https://www.who.int/health-topics/cancer},
  urldate={2025-04-30},
  organization={World Health Organization},
  note={Available at: \url{https://www.who.int/health-topics/cancer}}
}

@article{dagogo2018tumour,
  title={Tumour heterogeneity and resistance to cancer therapies},
  author={Dagogo-Jack, Ibiayi and Shaw, Alice T},
  journal={Nature reviews Clinical oncology},
  volume={15},
  number={2},
  pages={81--94},
  year={2018},
  publisher={Nature Publishing Group UK London}
}

@article{simpson2025challenges,
  title={Challenges of small cell lung cancer heterogeneity and phenotypic plasticity},
  author={Simpson, Kathryn L and Rothwell, Dominic G and Blackhall, Fiona and Dive, Caroline},
  journal={Nature Reviews Cancer},
  pages={1--16},
  year={2025},
  publisher={Nature Publishing Group UK London}
}

@article{hodel2018impact,
  title={The impact of biopsy sampling errors and the quality of surgical margins on local recurrence and survival in chondrosarcoma},
  author={Hodel, Sandro and Laux, Christoph and Farei-Campagna, Jan and G{\"o}tschi, Tobias and Bode-Lesniewska, Beata and M{\"u}ller, Daniel Andreas},
  journal={Cancer management and research},
  pages={3765--3771},
  year={2018},
  publisher={Taylor \& Francis}
}

@article{yang2022personalized,
  title={Personalized Biopsy Schedules Using an Interval-censored Cause-specific Joint Model},
  author={Yang, Zhenwei and Rizopoulos, Dimitris and Heijnsdijk, Eveline AM and Newcomb, Lisa F and Erler, Nicole S},
  journal={arXiv preprint arXiv:2209.00105},
  year={2022}
}

@article{duan2021evaluation,
  title={Evaluation and comparison of multi-omics data integration methods for cancer subtyping},
  author={Duan, Ran and Gao, Lin and Gao, Yong and Hu, Yuxuan and Xu, Han and Huang, Mingfeng and Song, Kuo and Wang, Hongda and Dong, Yongqiang and Jiang, Chaoqun and others},
  journal={PLoS computational biology},
  volume={17},
  number={8},
  pages={e1009224},
  year={2021},
  publisher={Public Library of Science San Francisco, CA USA}
}

@article{zhao2019molecular,
  title={Molecular subtyping of cancer: current status and moving toward clinical applications},
  author={Zhao, Lan and Lee, Victor HF and Ng, Michael K and Yan, Hong and Bijlsma, Maarten F},
  journal={Briefings in bioinformatics},
  volume={20},
  number={2},
  pages={572--584},
  year={2019},
  publisher={Oxford University Press}
}

@article{wang2023precision,
  title={Precision medicine: disease subtyping and tailored treatment},
  author={Wang, Richard C and Wang, Zhixiang},
  journal={Cancers},
  volume={15},
  number={15},
  pages={3837},
  year={2023},
  publisher={Multidisciplinary Digital Publishing Institute}
}

@article{liverani2021clustering,
  title={Clustering method for censored and collinear survival data},
  author={Liverani, Silvia and Leigh, Lucy and Hudson, Irene L and Byles, Julie E},
  journal={Computational Statistics},
  volume={36},
  pages={35--60},
  year={2021},
  publisher={Springer}
}

@article{bair2004semi,
  title={Semi-supervised methods to predict patient survival from gene expression data},
  author={Bair, Eric and Tibshirani, Robert},
  journal={PLoS biology},
  volume={2},
  number={4},
  pages={e108},
  year={2004},
  publisher={Public Library of Science San Francisco, USA}
}

@inproceedings{manduchideep,
  title={A Deep Variational Approach to Clustering Survival Data},
  author={Manduchi, Laura and Marcinkevi{\v{c}}s, Ri{\v{c}}ards and Massi, Michela C and Weikert, Thomas and Sauter, Alexander and Gotta, Verena and M{\"u}ller, Timothy and Vasella, Flavio and Neidert, Marian C and Pfister, Marc and others},
  booktitle={International Conference on Learning Representations}
}

@article{liu2020supervised,
  title={Supervised graph clustering for cancer subtyping based on survival analysis and integration of multi-omic tumor data},
  author={Liu, Cheng and Cao, Wenming and Wu, Si and Shen, Wenjun and Jiang, Dazhi and Yu, Zhiwen and Wong, Hau-San},
  journal={IEEE/ACM Transactions on Computational Biology and Bioinformatics},
  volume={19},
  number={2},
  pages={1193--1202},
  year={2020},
  publisher={IEEE}
}

@article{putter2007tutorial,
  title={Tutorial in biostatistics: competing risks and multi-state models},
  author={Putter, Hein and Fiocco, Marta and Geskus, Ronald B},
  journal={Statistics in medicine},
  volume={26},
  number={11},
  pages={2389--2430},
  year={2007},
  publisher={Wiley Online Library}
}

@inproceedings{guo2015robust,
  title={Robust subspace segmentation by simultaneously learning data representations and their affinity matrix.},
  author={Guo, Xiaojie},
  booktitle={IJCAI},
  pages={3547--3553},
  year={2015}
}

@article{zhou2022colorectal,
  title={Colorectal liver metastasis: molecular mechanism and interventional therapy},
  author={Zhou, Hui and Liu, Zhongtao and Wang, Yongxiang and Wen, Xiaoyong and Amador, Eric H and Yuan, Liqin and Ran, Xin and Xiong, Li and Ran, Yuping and Chen, Wei and others},
  journal={Signal transduction and targeted therapy},
  volume={7},
  number={1},
  pages={70},
  year={2022},
  publisher={Nature Publishing Group UK London}
}

@article{chow2019colorectal,
  title={Colorectal liver metastases: An update on multidisciplinary approach},
  author={Chow, Felix Che-Lok and Chok, Kenneth Siu-Ho},
  journal={World journal of hepatology},
  volume={11},
  number={2},
  pages={150},
  year={2019}
}

@article{fiz2020radiomics,
  title={Radiomics of liver metastases: a systematic review},
  author={Fiz, Francesco and Vigan{\`o}, Luca and Gennaro, Nicol{\`o} and Costa, Guido and La Bella, Ludovico and Boichuk, Alexandra and Cavinato, Lara and Sollini, Martina and Politi, Letterio S and Chiti, Arturo and others},
  journal={Cancers},
  volume={12},
  number={10},
  pages={2881},
  year={2020},
  publisher={MDPI}
}

@article{baghdadi2022imaging,
  title={Imaging of colorectal liver metastasis},
  author={Baghdadi, Azarakhsh and Mirpour, Sahar and Ghadimi, Maryam and Motaghi, Mina and Hazhirkarzar, Bita and Pawlik, Timothy M and Kamel, Ihab R},
  journal={Journal of Gastrointestinal Surgery},
  volume={26},
  number={1},
  pages={245--257},
  year={2022},
  publisher={Elsevier}
}

@article{nioche2018lifex,
  title={LIFEx: a freeware for radiomic feature calculation in multimodality imaging to accelerate advances in the characterization of tumor heterogeneity},
  author={Nioche, Christophe and Orlhac, Fanny and Boughdad, Sarah and Reuz{\'e}, Sylvain and Goya-Outi, Jessica and Robert, Charlotte and Pellot-Barakat, Claire and Soussan, Michael and Frouin, Fr{\'e}d{\'e}rique and Buvat, Ir{\`e}ne},
  journal={Cancer research},
  volume={78},
  number={16},
  pages={4786--4789},
  year={2018},
  publisher={American Association for Cancer Research}
}

@inproceedings{waldemar2022word2vec,
  title={Word2Vec embeddings for categorical values in synthetic tabular generation},
  author={Waldemar, Hahn and Martin, Sedlmayr and Markus, Wolfien},
  booktitle={2022 International Conference on Computational Science and Computational Intelligence (CSCI)},
  pages={613--622},
  year={2022},
  organization={IEEE}
}

@article{marin2010weighting,
  title={Weighting by inverse variance or by sample size in random-effects meta-analysis},
  author={Mar{\'\i}n-Mart{\'\i}nez, Fernando and S{\'a}nchez-Meca, Julio},
  journal={Educational and Psychological Measurement},
  volume={70},
  number={1},
  pages={56--73},
  year={2010},
  publisher={SAGE Publications}
}

@inproceedings{fisher2024inverse,
  title={Inverse-variance weighting for estimation of heterogeneous treatment effects},
  author={Fisher, A.},
  booktitle={Forty-first International Conference on Machine Learning},
  year={2024},
  month={July}
}

@article{lin2021combining,
  title={Combining the strengths of inverse-variance weighting and Egger regression in Mendelian randomization using a mixture of regressions model},
  author={Lin, Zhaoguan and Deng, Yusha and Pan, Wei},
  journal={PLoS Genetics},
  volume={17},
  number={11},
  pages={e1009922},
  year={2021},
  publisher={Public Library of Science}
}

@article{scheike2007direct,
  title={Direct modelling of regression effects for transition probabilities in multistate models},
  author={Scheike, Thomas H and Zhang, Mei-Jie},
  journal={Scandinavian Journal of Statistics},
  volume={34},
  number={1},
  pages={17--32},
  year={2007},
  publisher={Wiley Online Library}
}

@book{chung1997spectral,
  title={Spectral graph theory},
  author={Chung, Fan R. K.},
  volume={92},
  year={1997},
  publisher={American Mathematical Society}
}

@inproceedings{nie2014clustering,
  title={Clustering and projected clustering with adaptive neighbors},
  author={Nie, Feiping and Wang, Xiaoqian and Huang, Heng},
  booktitle={Proceedings of the 20th ACM SIGKDD international conference on Knowledge discovery and data mining},
  pages={977--986},
  year={2014}
}

@article{greenland1987quantitative,
  title={Quantitative methods in the review of epidemiologic literature},
  author={Greenland, Sander},
  journal={Epidemiologic Reviews},
  volume={9},
  pages={1--30},
  year={1987},
  publisher={Oxford University Press}
}

@book{hastie2009elements,
  title={The Elements of Statistical Learning: Data Mining, Inference, and Prediction},
  author={Hastie, Trevor and Tibshirani, Robert and Friedman, Jerome},
  year={2009},
  publisher={Springer Science \& Business Media},
  edition={2nd},
  series={Springer Series in Statistics}
}

@book{carlin2008bayesian,
  title={Bayesian Methods for Data Analysis},
  author={Carlin, Bradley P. and Louis, Thomas A.},
  year={2008},
  publisher={Chapman and Hall/CRC},
  edition={3rd},
  address={Boca Raton}
}

@book{andersen1993statistical,
  title={Statistical Models Based on Counting Processes},
  author={Andersen, Per K. and Borgan, {\O}rnulf and Gill, Richard D. and Keiding, Niels},
  year={1993},
  publisher={Springer-Verlag},
  address={New York},
  series={Springer Series in Statistics}
}

@inproceedings{zhou2004learning,
  title={Learning with local and global consistency},
  author={Zhou, Dengyong and Bousquet, Olivier and Lal, Thomas Navin and Weston, Jason and Sch{\"o}lkopf, Bernhard},
  booktitle={Advances in neural information processing systems},
  volume={16},
  pages={321--328},
  year={2004}
}

@article{belkin2003laplacian,
  title={Laplacian eigenmaps for dimensionality reduction and data representation},
  author={Belkin, Mikhail and Niyogi, Partha},
  journal={Neural computation},
  volume={15},
  number={6},
  pages={1373--1396},
  year={2003},
  publisher={MIT Press}
}

@inproceedings{wang2009manifold,
  title={Manifold alignment using Procrustes analysis},
  author={Wang, Chang and Mahadevan, Sridhar},
  booktitle={Proceedings of the 25th international conference on Machine learning},
  pages={1120--1127},
  year={2009}
}

@inproceedings{wang2020graph,
  title={Graph-regularized Cox model for survival prediction},
  author={Wang, Tianyi and Liu, Zhe and Jin, Yao and Xu, Jing and Yang, Yunan},
  booktitle={Proceedings of the AAAI Conference on Artificial Intelligence},
  volume={34},
  number={01},
  pages={619--626},
  year={2020}
}

@article{zwanenburg2016image,
  title={Image biomarker standardisation initiative},
  author={Zwanenburg, Alex and Leger, Stefan and Valli{\`e}res, Martin and L{\"o}ck, Steffen},
  journal={arXiv preprint arXiv:1612.07003},
  year={2016}
}

\end{document}